\documentclass[reprint,
 amsmath,amssymb,
 aps,
longbibliography]{revtex4-2}
\usepackage{graphicx}% Include figure files
\usepackage{dcolumn}% Align table columns on decimal point
\usepackage{bm}% bold math
\usepackage[caption=false]{subfig}
\usepackage{hyperref}
\newcommand{\V}[1]{\textup{#1}}

\begin{document}
\onecolumngrid
This work has been published in A. Martí, E. Antolín, and I. Ramiro, “Thermodynamics of the Monoenergetic Energy-Selective Contacts of a Hot-Carrier Solar Cell,” Phys. Rev. Appl., vol. 18, no. 6, p. 064048, Dec. 2022, doi: \href{https://doi.org/10.1103/PhysRevApplied.18.064048}{10.1103/PhysRevApplied.18.064048}. The American Physical Society has the copyright of the published version.
\preprint{APS/123-QED}

\title{Thermodynamics of the\\ mono-energetic energy selective Contacts \\ of the hot carrier solar cell}% Force line breaks with \\

\author{Antonio Mart\'i}
 \email{antonio.marti@upm.es}
 \author{Elisa Antol\'in}
\author{I\~{n}igo Ramiro}%
\altaffiliation[Now at ] {Universidade Nova de Lisboa}

\affiliation{%
 Instituto de Energ\'ia Solar - Universidad Polit\'ecnica de Madrid\\
 ETSI Telecomunicaci\'on, Ciudad Universitaria, 28040 Madrid, Spain, EU % \textbackslash\textbackslash
}%



\begin{abstract}
The hot carrier solar cell (HCSC) has the potential for converting solar energy into electrochemical energy with an efficiency of 85.4\%. For this, in addition to an idealized light absorber,  the HCSC has to be connected to the external load by means of the so-called \emph{mono-energetic energy selective contacts} (ESCs). However, the thermodynamic properties that these types of contact have to exhibit, such as their electric, thermal conductivity and Seebeck coefficient, have not been explored. This paper aims to fill this gap. In this respect, we model electron transport in non-ideal ESCs using the transport theory proposed by Datta and Landauer which has allowed us to calculate the value of these parameters as a function of the temperature and electrochemical potential of operation.  Our findings also reveal  that, to preserve the HCSC efficiency above 82\%, the ESCs could require in the order of $3 \times 10^{19}$ cm$^{-3}$ electron states. As the ESCs depart from ideality, the temperature of the hot carriers at which optimum efficiency is obtained increases to above 2540 K.  The mono-enenergetic selective contact characterized by the highest energy demands an electric, thermal conductivity and Seebeck coefficient  that, when combined, are characterized by a high thermoelectric figure of merit $(ZT\approx 8)$. We are not aware of any material exhibiting this figure of merit which illustrates the difficulty in putting the HCSC concept into practice. Conversely, our work supports the idea that pursuing materials capable of transporting electrons ballistically through mono-energetic electron channels can provide the key for achieving materials characterized by high $ZT$. 
%\begin{description}
%\end{description}
\end{abstract}

\maketitle

%\tableofcontents

\section{\label{sec:level1}Introduction\protect %\\ The line
%break was forced \lowercase{via} \textbackslash\textbackslash
}

The initial formulation of  the hot carrier solar cell concept (HCSC) was proposed by Ross and Nozik in 1982  \cite{RossNozik}. It consisted of the idea that extracting photogenerated carriers from a solar cell, before giving them time to lose part of this energy through thermalization, would lead to a solar cell with a limiting efficiency of 85.4\%. Later on, Würfel \cite{WURFEL199743} realized that, in addition, the hot carriers generated should be prevented from increasing their entropy while being transported through the contacts, from the hot side at the absorber, to the cold side at the terminals and vice-versa. We shall assume, without loss of generality in our treatment, that these \emph{carriers} are \emph{electrons}. Proceeding in this way, during their transport through the contacts, electrons would increase their electrochemical energy while cooling down leading to the appearance of the output voltage of the cell. The contacts capable of achieving this transformation were designated as \emph{mono-energetic energy selective contacts} (ESCs). These were conceived as an electron transport channel in which electrons were allowed to have only one single energy.  

In addition  to the original aforementioned references,   Knig et al. \cite{knig_hot_2010}, Köning et al. \cite{konig_non-equilibrium_2020} and Zhang et al. \cite{zhang_review_2021} have extensively reviewed material candidates for the implementation of HCSCs. Dimmock et al. \cite{dimmock_demonstration_2014, dimmock_metallic_2019} have reported a hot carrier signature in the current-voltage characteristics of devices implemented with GaAs and thin metal absorbers followed by resonant tunneling structures as ESCs.
Nguyenet al. \cite{nguyen_quantitative_2018} have pointed at hot carrier dynamics as the mechanism behind the current and voltage enhancement in InGaAsP quantum well solar cells. Li et al. \cite{li_slow_2017} have suggested the use of colloidal halide perovskite nanocrystals to overcome the limitations of inorganic materials for slowing down carrier cooling. Esmaielpour et al.\cite{esmaielpour_determination_2020} have applied a contactless measurement using photoluminiscence to investigate the Seebeck coefficient of InGaAs multi-quantum well structures for its possible application as an ESC for an HCSC. Williams at al. \cite{williams_hot_2013} have discussed the use of graphene nanomaterials as absorber material candidates. 
 Additionally, the interested reader can consult our works in \cite{Marti2019,marti_ele_2013} for a review of the role of electron electrochemical potential(s) in the operation of the HCSC. 

In spite of these advances, we think a systematic study of the transport properties associated to the ESCs in terms of electric and thermal conductivity as well as their thermoelectric properties is still missing. This work aims to fill this gap. To do so, in  section II we shall revisit the general structure of an ideal HCSC consisting of the absorber and the ESCs. In section III, we shall focus on a detailed description of the ESCs. In sections IV and V we review, respectively, the conventional theory that allows the limiting efficiency of an HCSC to be calculated, under the assumption of ideal ESCs, and its current-voltage characteristic. In section VI, we shall consider non-ideal ESCs and recalculate the limiting efficiency of the HCSC as a function of the \emph{reverse saturation particle current density}, $\mathit{\Gamma}$, of the ESC. Finally, in section VII we calculate the electric and thermal conductivity of the ESC, its Seebeck coefficient and $ZT$ figure of merit. The model we use is described in detail in the Appendix.

\section{\label{sec:general} General description of a hot carrier solar cell \protect}

\begin{figure}[htp] 
\subfloat{%
  \includegraphics[clip,width=\columnwidth]{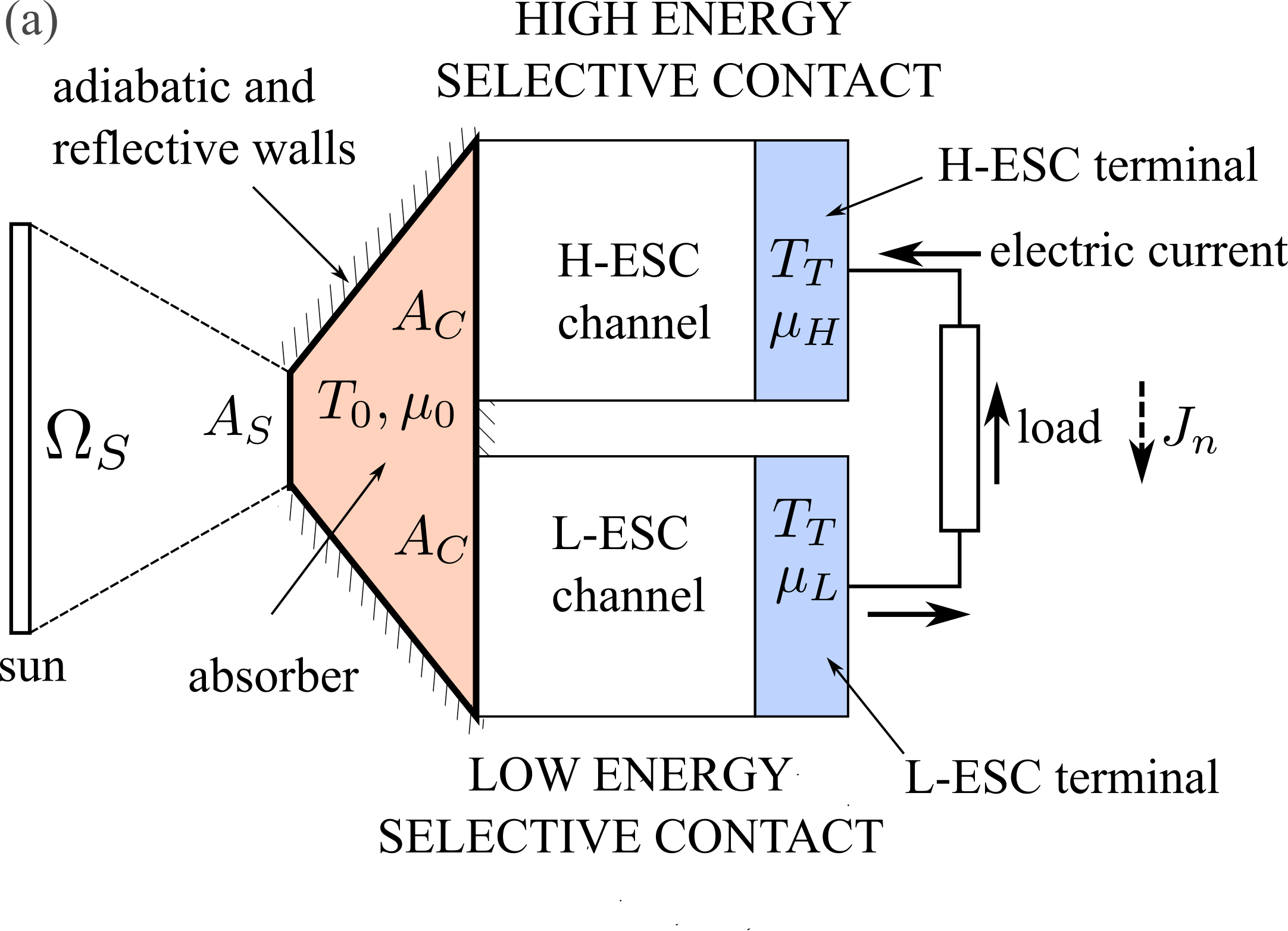}%
}

\subfloat{%
  \includegraphics[clip,width=0.8\columnwidth]{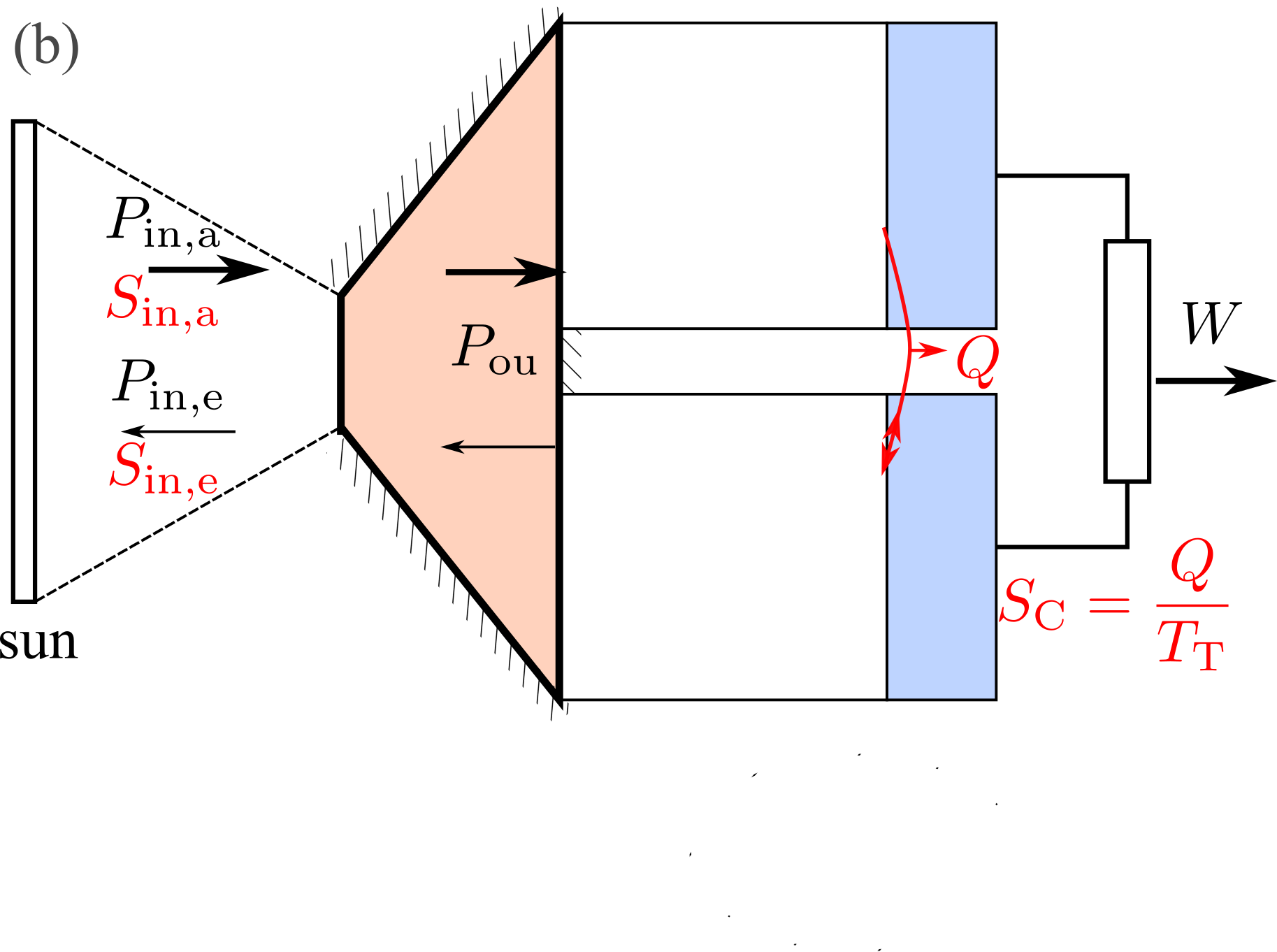}%   
}
\caption{\label{fig:HCSC_illumination} (a) Scheme showing the illumination and contacting scheme of an HCSC. The solid arrow indicates the direction of the electric current. The dashed arrow indicates the direction of the particle (electrons) current. The operation of the absorber is  characterized by the temperature $T_0$ and electrochemical potential $\mu_0$. The absorber is assumed to absorb all the sunlight and have no bandgap. The terminal of the high energy selective contact (H-ESC) operates at temperature $T_\V{T}$ and electrochemical potential $\mu_\V{T}=\mu_\V{H}$. The terminal of the low energy selective contact (L-ESC)  operates at temperature $T_\V{T}$ and electrochemical potential $\mu_\V{L}$; (b) The same scheme but illustrating the power inputs and outputs ($P$), heat released per unit of time ($Q$), electric power ($W$) and entropy rates associated ($S$).}
\end{figure} 
Fig. \ref{fig:HCSC_illumination} sets out the basic structure of an HCSC. It consists of two parts: the light absorber and two ESCs. For simplicity, the absorber will be assumed  to be illuminated by the sun, modeled as a blackbody at the temperature $T_\V{S}=$ 6000~K. Photons are received from the sun with an étendue $\mathit{\Omega}_\V{S}$ \cite{WELFORD19899} and the absorber emits photons back to the sun with the same étendue. This approach is equivalent to assuming that the HCSC operates at maximum light concentration ($X_\mathrm{S}=46050$, for the area of the absorber receiving sunlight being surrounded by air). 	$X_\mathrm{S}$ will often be used as a scaling factor in the results that follow in order to obtain human readable values throughout this work. 

The absorber can be visualized as an electron gas that, heated by the sun, in steady-state operation, and depending on the load, it reaches the temperature $T_0$. In addition, electrons in the absorber all share the same electrochemical potential, $\mu_0$. The existence of a single electrochemical potential for electrons in the absorber was reasoned by Würfel in the so-called \emph{energy conservation model} \cite{WURFEL199743} for the HCSC, that will be used throughout this work, and contrasts with the \emph{ particle conservation model}, in which two electrochemical potentials (for electrons and holes) were assumed instead, and which is known to lead to non-physically sound results \cite{Wurfel2005}. The existence of a single electrochemical potential $\mu_0$ arises from the hypothesis that electrons only interact with photons and among themselves, and not with lattice phonons. In \cite{WURFEL199743} it was also shown that considering a hot temperature for these electrons and a single electrochemical potential was equivalent to considering electrons were cold but their population at each energy state in the absorber was described by its own electrochemical potential. In addition to \cite{WURFEL199743}, the reader can consult \cite{Marti2019, marti_ele_2013} for a detailed discussion on this subject. The absorber has no bandgap because it is also known that zero gap is the option that leads to the maximum efficiency. 

On the other hand, we have the two aforementioned ESCs, one for electrons to exit the absorber, named \emph{high energy selective contact} (H-ESC) and another one, named \emph{low energy selective contact} (L-ESC), to allow electrons to return to the absorber. The ESCs connect, through their respective electron mono-energetic \emph{channels}, the absorber with the terminals, which are in contact with the load where the electric power is extracted. Both terminals are at room temperature, $T_\V{T}=$ 300 K but, as a consequence of the energy conversion process, the terminal of the H-ESC is characterized by the electrochemical potential $\mu_\V{H}$ and, the terminal of the L-ESC, by the electrochemical potential, $\mu_\V{L}$. Since these ESCs will focus a great deal of the attention in our work, they will be described with more detail in the next section. 

Note that, due to the negative charge of the electrons, the electrical current (units A\kern 0.15emcm$^{-2}$) flows (solid arrow) in the opposite direction to the movement of the electrons considered as particles. To produce positive electric power at the load, electrons have to exit from the H-ESC and return via the L-ESC. In order to avoid confusion, in this work we shall always deal in terms of particle current density, $J_\V{n}$, whose units are \emph{number of electrons per unit of time and unit of area}. This particle current density can then be easily transformed, when required, into electrical current density (with units of \emph{amperes} per \emph{unit of area}) by multiplying by $-e$, $e$ being the electron charge in absolute value.

For major generality, the area in which the absorber receives photons from the sun, $A_\V{S}$, will be considered different from the area of the contacts, $A_\V{C}$. Also, for simplicity, the area of both ESCs will be assumed to be the same and uniform all along the contact. This possible difference between areas $A_\V{S}$ and $A_\V{C}$  had not been previously considered to our knowledge in the context of HCSCs. In this respect, we shall define $X$, the ratio between both areas, as:
\begin{eqnarray}\label{eq:X}
X=\frac{A_\V{C}}{A_\V{S}}
\end{eqnarray}
The remaining surface surrounding the absorber can be assumed, either negligible, or covered by ideal adiabatic reflecting walls so that neither heat nor light are allowed to escape through this surface.

\begin{figure}[htp] 
\subfloat{%
  \includegraphics[clip,width=\columnwidth]{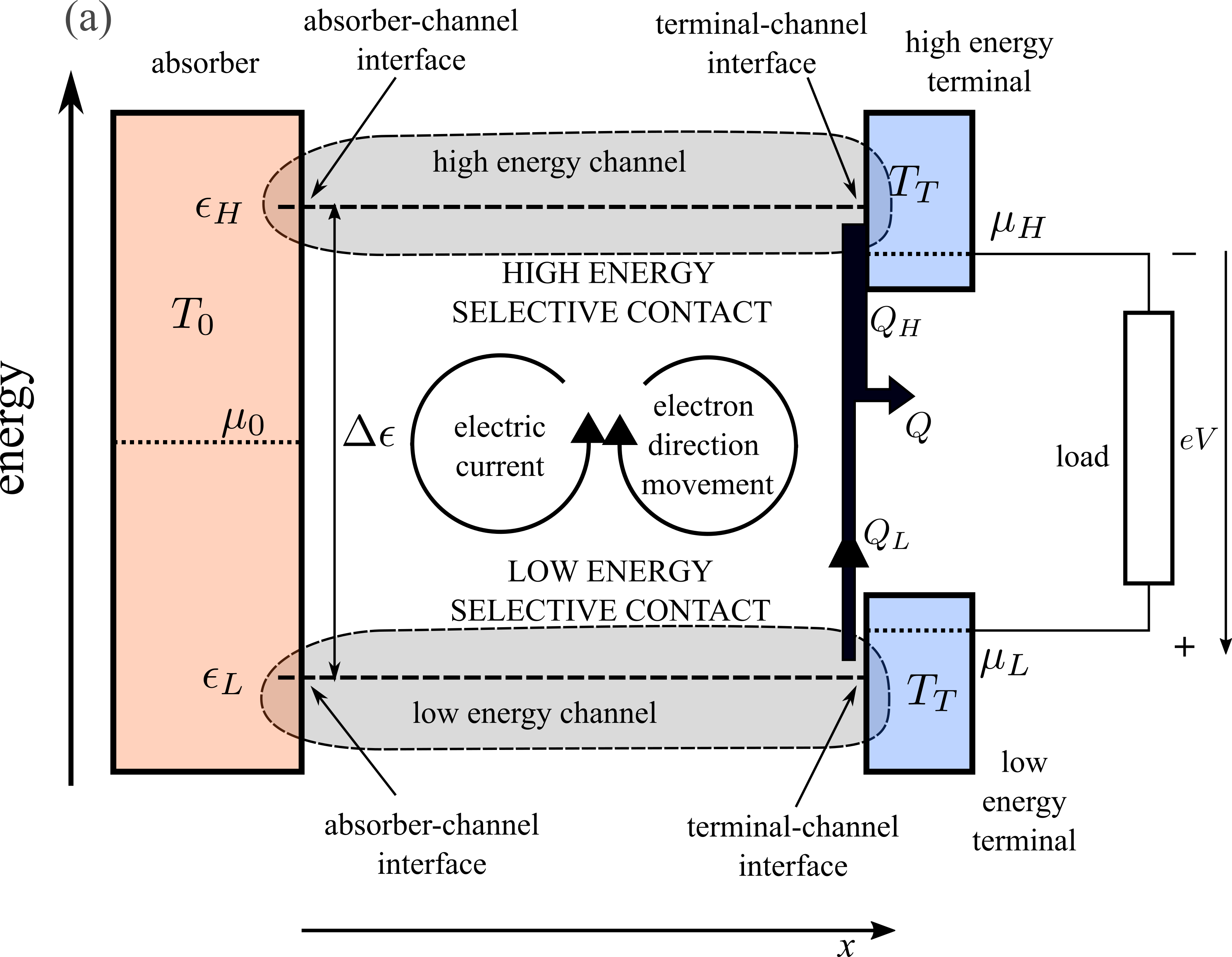}%
}

\subfloat{%
  \includegraphics[clip,width=\columnwidth]{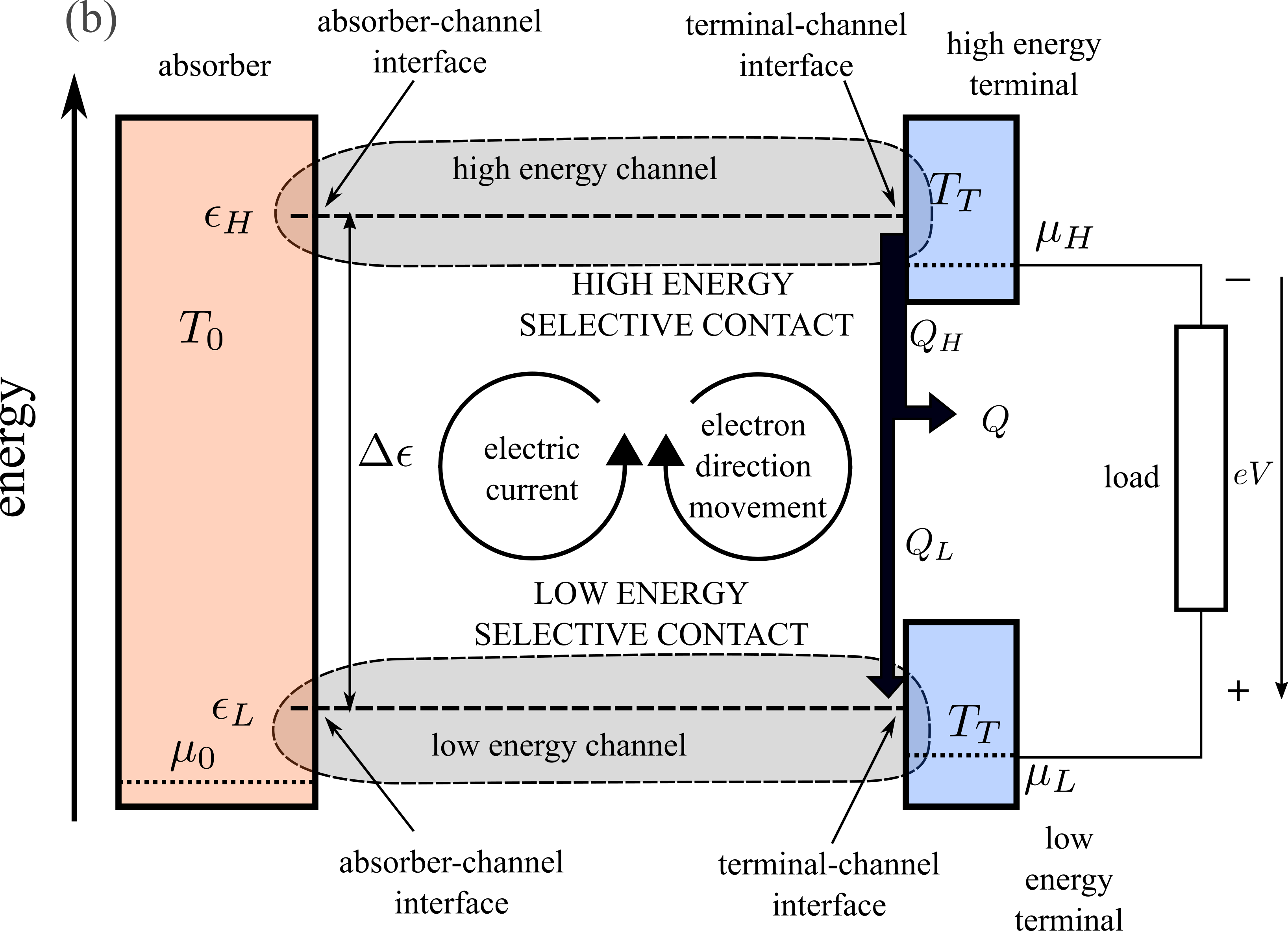}%   
}

\caption{\label{fig:HCSC} Two different schemes illustrating the operation of a hot carrier solar cell and the relative position of the energy of the selective contacts with respect to the electrochemical potentials: (a)  $\epsilon_\V{H}>\mu_\V{H}>\mu_0$ and $\epsilon_\V{L}<\mu_\V{L}<\mu_0$;  (b) $\epsilon_\V{H}>\mu_\V{H}>\mu_0$ and $\epsilon_\V{L}>\mu_\V{L}>\mu_0$.}

\end{figure}

\section{Detailed Description of the energy selective contacts \protect} \label{sec:ESC}

Each ESC will be considered  to consist of three parts: the \emph{channel} through which the electrons are transported, the \emph{absorber-channel interface} and the \emph{channel-terminal interface}. In the channels, only electrons with energy $\epsilon$ can exist, this being the reason why they are named mono-energetic. In this respect, in the H-ESC,   $\epsilon=\epsilon_\V{H}$ and in the L-ESC channel, $\epsilon=\epsilon_\V{L}$ (Fig. \ref{fig:HCSC}).

In the HCSC energy conservation model \cite{WURFEL199743}, in order to calculate the maximum conversion efficiency, it is assumed that the ESCs transport electrons isentropically. Following this energy conservation model, this condition is claimed to be satisfied when:
\begin{eqnarray} {\label{eq:muH}}
\frac{\epsilon_\V{H}-\mu_0 }{k_\V{B} T_0}=\frac{\epsilon_\V{H}-\mu_\V{H}}{k_\V{B} T_\V{T}}\Longrightarrow \nonumber \\
\mu_\V{H}-\mu_0=\left(\epsilon_\V{H}-\mu_0 \right) \left( 1-\frac{T_\V{T}}{T_0} \right)
\end{eqnarray}
\begin{eqnarray}{\label{eq:muL}}
\frac{\epsilon_\V{L}-\mu_0 }{k_\V{B} T_0}=\frac{\epsilon_\V{L}-\mu_\V{L}}{k_\V{B} T_\V{T}}\Longrightarrow \nonumber \\
\mu_\V{L}-\mu_0=\left(\epsilon_\V{L}-\mu_0 \right) \left( 1-\frac{T_\V{T}}{T_0} \right)
\end{eqnarray}
for the high and low ESCs respectively. Note that, since $0<(1-T_\V{T}/T_\V{0})<1$, $|\mu_\V{H}-\mu_0|<|\epsilon_\V{H}-\mu_0|$ and $|\mu_\V{L}-\mu_0|<|\epsilon_\V{L}-\mu_0|$. In addition, in order to preserve the sign at both sides of Eq.\ref{eq:muH} , if we assume, as we have done in Fig.  \ref{fig:HCSC}, that $\epsilon_\V{H}>\mu_0$, then we must also have $\mu_\V{H}>\mu_0$. Concerning the energy of the L-ESC, $\epsilon_\V{L}$, we shall consider for illustrative purposes two possible cases. In the first, illustrated in Fig.  \ref{fig:HCSC}a, we shall consider $\epsilon_\V{L}<\mu_0$ what, with the considerations above, leads us to $\mu_0>\mu_\V{L}>\epsilon_\V{L}$. In the second case,  illustrated in Fig.  \ref{fig:HCSC}b, we shall consider  $\epsilon_\V{L}>\mu_0$ what, also with the considerations above, lead us to $\epsilon_\V{L}>\mu_\V{L}>\mu_0$.

Let us now first focus our attention in the H-ESC as it has been sketched  in both Fig. \ref{fig:HCSC}a and b and think now of this contact as being implemented by means of an actual material.  We observe that, as we move away from the absorber towards the  terminal, the electrochemical potential of the electrons rises from $\mu_0$ to $\mu_\V{H}$, that is, we expect a positive electron electrochemical gradient at this contact:
\begin{eqnarray}
\frac{d\mu}{dx}>0
\end{eqnarray}
However, if this were the only existing driving force, the electrons would move from right to left whereas, in our illustration, we have assumed they travel from left to right.  Another electron driving force at the H-ESC is then necessary for preserving the coherence in the direction of the electron movement in this contact. This other driving force is the inverse temperature gradient, $1/T$ so that:
\begin{eqnarray}
\frac{d(1/T)}{dx}>0
\end{eqnarray}
In the linear approximation, the particle current density, $J_\V{n}$, is given by \cite{Callen1}:
\begin{eqnarray}\label{eq:Jn}
J_\V{n}=-L_{1,1}\frac{1}{T}\frac{d\mu}{dx}-L_{1,2}\frac{d(1/T)}{dx}
\end{eqnarray}
where we have assumed a one dimensional model extending in the direction $x$ for the ESC. The coefficients $L_{1,1}$ and $L_{1,2}$ are related to the electric conductivity $\sigma$ and Seebeck coefficient, $\xi$, by:
\begin{eqnarray}
L_{1,1}=\frac{\sigma T}{e^2 }\\
L_{1,2}=\frac{\xi \sigma T^2}{e}
\end{eqnarray}
and where $\xi<0$ for electrons and $\xi>0$ for holes.  Since at the H-ESC we must have $J_\V{n}>0$, given that $L_{1,1}>0$, $d\mu/dx>0$ and $d(1/T)/dx>0$ then, according to Eq. \ref{eq:Jn} we must have that:
\begin{eqnarray}
-L_{1,2} >0 
\end{eqnarray}
which implies that $\xi<0$ and, therefore, that the current across the H-ESC is transported by electrons.

We shall analyze now the L-ESC. At this contact, we must have $J_\V{n}<0$. On the other hand,  we also have that $d(1/T)/dx>0$ for both the case illustrated in Fig. \ref{fig:HCSC}a as in \ref{fig:HCSC}b. However we have $d\mu/dx<0$ for the case in Fig. \ref{fig:HCSC}a, what lead us to the fact that we must have $L_{1,2}>0$, that is, to an electric current across the L-ESC necessarily consisting of holes ($\xi>0$). Notice this is not in contradiction with the general scheme represented in Fig. \ref{fig:HCSC}a for the direction of the electric current since the hole current carries positive charge in the direction opposite to the electron movement. On the other hand, since we have $d\mu/dx<0$ for the case in Fig. \ref{fig:HCSC}b, $L_{1,2}$ could be either positive or negative an still allow for $J_\V{n}<0$.

Another relevant observation refers to the heat current density, $J_\V{q}$, transported by the electrons along the contacts. In the linear approximation, this current density is related to the electrochemical potential, $\mu$, and temperature gradient, $1/T$, by \citep{Callen1}:
\begin{eqnarray}\label{eq:JQ}
J_\V{q}=L_{2,1}\frac{1}{T}\frac{d\mu}{dx}+L_{2,2} \frac{d(1/T)}{dx}
\end{eqnarray}
The coefficient $L_{2,2}$ is related to the contact thermal conductivity $\kappa$, Seebeck coefficient and electric conductivity by \citep{Callen1}:
\begin{eqnarray}
L_{22}=T^2(\kappa+T\xi^2 \sigma)
\end{eqnarray}
which is positive. Therefore, the second term in Eq. \ref{eq:JQ} drives the heat away from the absorber towards the terminals both in the H-ESC as in the L-ESC. On the other hand, according to the Onsager relation \citep{Callen1}, we must have $L_{2,1}=L_{1,2}$. This implies that the first term in Eq. \ref{eq:JQ}, which is negative for the H-ESC since we obtained that $\xi<0$ in order to have $J_\V{n}>0$, brings the heat back to the absorber. This will also be the case for the L-ESC configuration in Fig. \ref{fig:HCSC}a, since the current across the L-ESC had to consist of holes ($L_{1,2}>0$ and $d\mu/dx<0$). However, in the case illustrated in \ref{fig:HCSC}b, the first term in  \ref{eq:JQ} will also drive heat away from the absorber through the L-ESC if electric current consists of holes.

Finally, Eqs. \ref{eq:muH} and \ref{eq:muL} have an interesting physical interpretation. Hence, taking $\mu_0$ as the reference of zero energy, in the H-ESC, each electron, takes the energy $\epsilon_\V{H}$ from the absorber and converts it into electrochemical work $\mu_H$ with the Carnot efficiency that is determined by the absorber temperature, $T_0$ (at the hot side) and the terminal temperature, $T_\V{T}$ (at the cold side).  The energy difference between $\epsilon_\V{H}$ and $\mu_\V{H}$ must be lost as heat $Q_\V{H}$ at the H-ESC contact. At the load, each electron releases an electrochemical energy $\mu_\V{H}-\mu_\V{L}$ and leave from the load with electrochemical energy $\mu_\V{L}$. Then,  Eq. \ref{eq:muL}, reveals that the L-ESC, in the case of Fig. \ref{fig:HCSC}b performs as a refrigerator that uses the electrochemical work $\mu_\V{L}$ and heat $Q_\V{L}$ to return electrons with energy $\epsilon_\V{L}$ to the absorber. In contrast, in the case of Fig. \ref{fig:HCSC}a, part of the electrochemical energy $\mu_\V{L}$ of the electrons that enter into the L-ESC from the terminal is dissipated as heat to produce electrons with energy $\epsilon_L$ that enter back into the absorber. In both cases, the net power released as heat at the terminal-channel interfaces, $Q=Q_\V{H}+Q_\V{L}$, for the case represented in Fig. \ref{fig:HCSC}a, and    $Q=Q_\V{H}-Q_\V{L}$ for the case represented in \ref{fig:HCSC}b, is given by: 
\begin{eqnarray}\label{eq:Q}
Q=eA_\V{C}J_\V{n}\left[(\epsilon_\V{H}-\mu_\V{H})-(\epsilon_\V{L}-\mu_\V{L}) \right]
\end{eqnarray}

\section{\label{sec:limiting} Limiting efficiency of a Hot Carrier Solar Cell for ideal energy selective contacts \protect }

The limiting efficiency of an HCSC can be obtained from the energy conservation model described in \cite{WURFEL199743}. According to this model, adapted here for the general case in which $A_\V{S} \neq A_\V{C}$, the net energy per unit of time entering into the absorber, $P_\V{in}$, has to be equal to the net power output, $P_\V{ou}$,
\begin{eqnarray}\label{eq:Pequals}
P_\V{in}=P_\V{ou}
\end{eqnarray}
$P_\V{in}$ corresponds to the difference between the power carried by the photons absorbed from the sun, $P_\V{in,a}$, and the power carried by the photons emitted back to the sun from the absorber, $P_\V{in,e}$ (Fig. \ref{fig:HCSC_illumination}b). Assuming maximum light concentration and both the sun and the absorber as black bodies characterized by the temperatures $T_\V{S}$ and $T_0$ respectively, we have: 
\begin{eqnarray}\label{eq:P_in}
P_\V{in,a}=A_\V{S} \hat{\sigma} T_\V{S}^4 \\
P_\V{in,e}=A_\V{S} \hat{\sigma} T_0^4 
\end{eqnarray}
$\hat{\sigma}$ being the Stefan-Boltzmann constant. Therefore,
\begin{eqnarray}\label{eq:P_in}
P_\V{in}=P_\V{in,a}-P_\V{in,e}=A_\V{S} \hat{\sigma} \left(T_\V{S}^4-T_0^4 \right)
\end{eqnarray}
The efficiency with which power from the sun is captured by the absorber, $\eta_\V{A}$, is given then by:
\begin{eqnarray}{\label{eq:etaA}}
\eta_\V{A}=\frac{P_\V{in,a}-P_\V{ou,e}}{P_\V{in,a}}=\left(1-\frac{T_0^4}{T_\V{S}^4} \right)
\end{eqnarray}

 On the other hand, since the only power being extracted takes place through the contacts,
\begin{eqnarray}\label{eq:P_ou}
P_\V{ou}=eA_\V{C} J_\V{n} \epsilon_\V{H}-eA_\V{C}J_\V{n} \epsilon_\V{L}= eA_\V{C} J_\V{n}  \Delta \epsilon
\end{eqnarray}
with
\begin{eqnarray}\label{eq:deltaEps}
\Delta\epsilon=\epsilon_\V{H}-\epsilon_\V{L}
\end{eqnarray}
The efficiency, $\eta_\V{B}$, with which the contacts convert the net power received from the absorber, $P_\V{ou}$, into electric power, $W$, where
\begin{eqnarray}
W=eA_\V{C}J_\V{n} (\mu_\V{H}-\mu_\V{L})
\end{eqnarray}
is then given by: 
\begin{eqnarray}\label{eq:etaB}
\eta_B=\frac{W}{P_\V{ou}}=\frac{\mu_\V{H}-\mu_\V{L}}{\epsilon_\V{H}-\epsilon_\V{L}}=\left(1-\frac{T_\V{T}}{T_0}\right)
\end{eqnarray}
where we have used Eqs. \ref{eq:muH} and \ref{eq:muL}. This result, combined with Eq. \ref{eq:etaA}, lead us to the well-known limiting efficiency of the HCSC, $\eta$:
\begin{eqnarray}
\eta=\eta_\V{A} \eta_\V{B}=\left(1-\frac{T_0^4}{T_\V{S}^4} \right)\left(1-\frac{T_\V{T}}{T_0}\right)=0.854
\end{eqnarray}
obtained for an optimized value $T_0=2544$ K. As expected, the difference $P_\V{ou}-W$ equals the power released as heat at the terminal-channel interface, $Q$, anticipated by Eq. \ref{eq:Q}.

\section{\label{sec:JV} \protect Current-voltage characteristic of a Hot carrier solar cell with ideal energy selective contacts }

The electrochemical potential difference $\mu_\V{H}-\mu_\V{L}$ is perceived at the terminals as the output voltage of the cell, $V$:
\begin{eqnarray}\label{eq:V}
\mu_\V{H}-\mu_\V{L}=e V
\end{eqnarray}

The reader must be aware that this equation, due to the negative charge of the electrons, is consistent with the fact that the voltage at the L-ESC terminal has to be higher than the voltage at the H-ESC terminal which is also consistent with the direction of the \emph{particle} current density.
 
Once the values of $\epsilon_\V{H}$ and $\epsilon_\V{L}$ are fixed for a given HCSC, then, for a given voltage $V$, substituting Eq. \ref{eq:V} into Eqs. \ref{eq:muH} and \ref{eq:muL}, allow us to obtain the absorber operation temperature $T_0$ that is given by:
\begin{eqnarray}
T_0=T_\V{T}\frac{\epsilon_\V{H}-\epsilon_\V{L}}{\epsilon_\V{H}-\epsilon_\V{L}-\mu_\V{H}+\mu_\V{L}}=T_\V{T}\frac{\Delta \epsilon}{\Delta \epsilon -e V}
\end{eqnarray}

\begin{figure}
\includegraphics[width=\columnwidth]{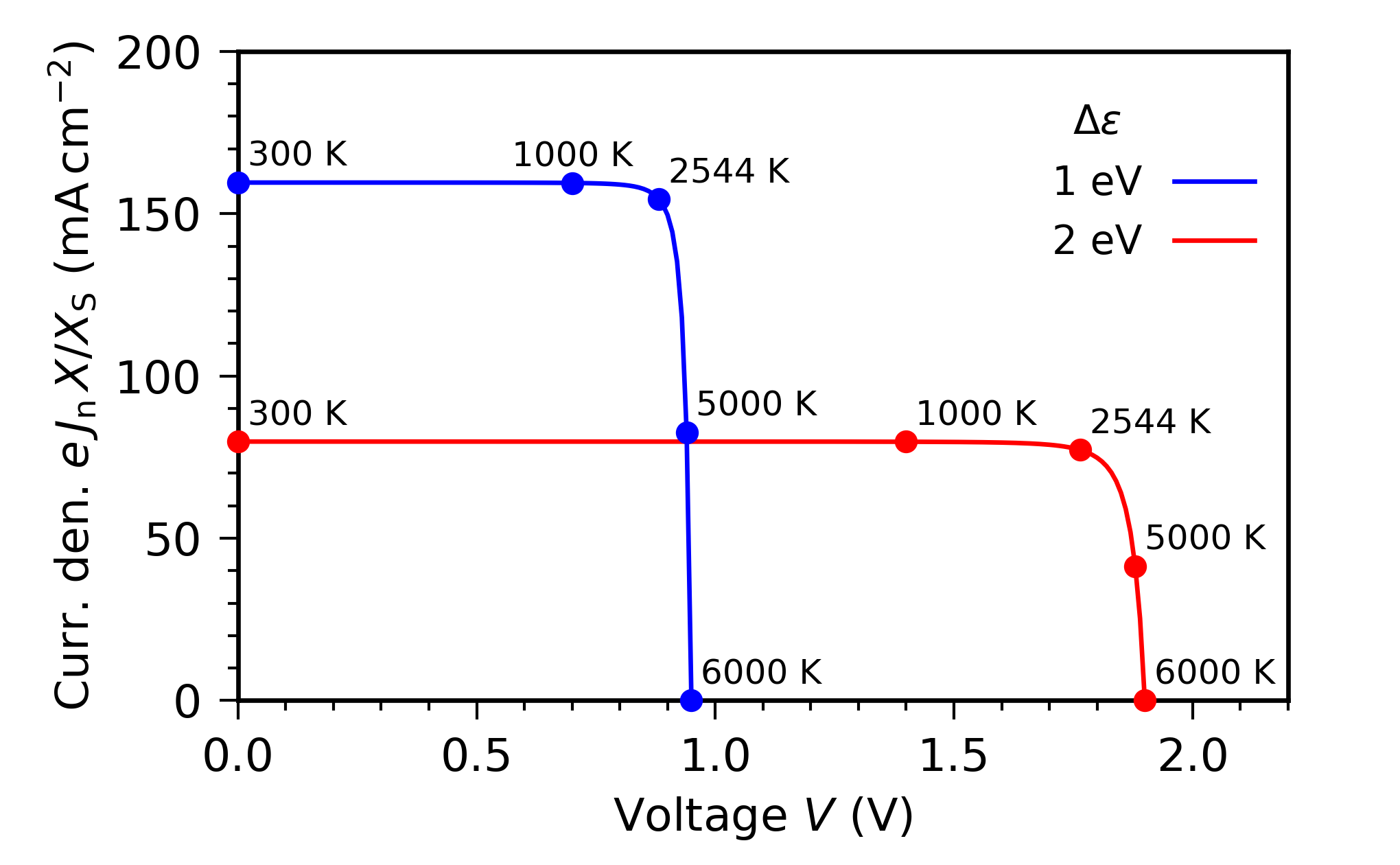}
\caption{\label{fig:JV} Two examples of current-voltage characteristis  of an ideal HCSC corresponding to $\Delta \epsilon = 1$ eV and $\Delta \epsilon =2$ eV. Several absorber temperatures are also indicated. The maximum power point always occurs at $T_0=2544$ K leading also always to an efficiency of 85.4 \%. Plots have been scaled by the maximum solar concentration factor $X_S$=46050. $X$ is the ratio between the contact and absorber area ($A_\V{C}/A_\V{S})$. }
\end{figure}

The temperature $T_0$ can now be inserted into Eq. \ref{eq:P_in} to obtain, together with Eqs. \ref{eq:Pequals} and \ref{eq:P_ou}, the electron current density $eJ_\V{n}$ circulating through the cell. Fig. \ref{fig:JV} plots the current density obtained so far for two examples: $\Delta \epsilon =$ 1~eV and $\Delta \epsilon =$ 2~eV. 

These examples illustrate that, although the shape of the current density voltage characteristic of an HCSC depends only on the choice of $\Delta \epsilon=\epsilon_\V{H}-\epsilon_\V{L}$, the maximum efficiency does not, always being equal to 85.4\% and always being  obtained for a temperature $T_0=2544$ K. In this respect, an HCSC with large $\Delta \epsilon$ will produce a solar cell with high open-circuit voltage and low short-circuit current. Conversely, an HCSC with low 
$\Delta \epsilon$ will produce a solar cell with low open-circuit voltage and large short-circuit current. Therefore, in an HCSC, $\Delta \epsilon$ plays a role similar to that of the semiconductor gap in a single-gap solar cell but without impacting the efficiency of the solar cell. It is also worth noticing that, the short-circuit current is always obtained for $T_0=300$ K and, the open-circuit voltage, for $T_0=6000$ K.

Although  transport through the ESCs is still being considered isentropic at this stage, the full converter (consisting of the absorber plus the ESCs) is not operating, in general, reversibly. In this respect, notice that the irreversible production rate at the converter, $S_\V{irr}$ is given by (Fig.  \ref{fig:HCSC_illumination}b) \cite{landsberg}:
\begin{eqnarray}
S_\V{irr}=S_\V{in,a}-S_\V{in,e}-S_\V{C}
\end{eqnarray}
where
\begin{eqnarray}
S_\V{in,a}=A_\V{S}\frac{4}{3} \hat{\sigma} T_\V{S}^3
\end{eqnarray}
is the entropy absorbed per unit of time associated to the photons absorbed from the sun, 
\begin{eqnarray}
S_\V{in,e}=A_\V{S}\frac{4}{3} \hat{\sigma} T_\V{0}^3
\end{eqnarray}
is the entropy leaving the converter per unit of time associated to the photons emitted from the absorber, and 
\begin{eqnarray}
S_\V{C}=\frac{Q}{T_\V{T}}
\end{eqnarray}
is the entropy associated to the release of heat $Q$ to the heat sink, assumed at ambient temperature which equals the temperature of the terminals, $T_\V{T}$

Full reversible operation of the solar converter would imply $S_\V{irr}=0$ which occurs when $T_0=T_S=6000 \rm{K}$. For this temperature, $S_\V{in,a}=S_\V{in,e}$ and, as it can be seen in the particle current density plots in  Fig. \ref{fig:JV}, $T_0=T_S=6000 \rm{K}$ also corresponds to the converter operating in open-circuit conditions in which no current is extracted from the cell $J_\V{n}=0$ and, therefore, according to Eq. \ref{eq:Q}, also $Q=0$.

For other absorber temperatures, such that $T_\V{0} \neq T_\V{S}$, the total energy conversion process is not reversible and current can be extracted from the converter even if transport through the ESCs is considered isentropic.  In an oversimplified comparison, we could make the analogy that we would be dealing with a solar cell (consisting, for example, of a pn semiconductor junction) in which the contacts are assumed ideal (no losses): the fact that the contacts are assumed without losses does not prevent current from the solar cell from being extracted but that this current is extracted more efficiently.

It is also relevant to note that the particle current density, $J_\V{n}$, appears in the plot in Fig. \ref{fig:JV} multiplied by $X/X_\V{S}$ and therefore, it can be made arbitrarily low by increasing $X$, that is, the ratio between the contact area, $A_\V{C}$, and the area with  which the absorber collects light, $A_\V{S}$ (Eq.\ref{eq:X}). 

\section{\protect  Current density vs voltage characteristic and limiting efficiency of a hot carrier solar cell with non-ideal ESCs} 
\label{sec:noiso}

In \cite{Datta}, S. Datta, building on previous models by Landauer \cite{landauer_1957,landauer_1992}, has developed a transport theory for a mono-energetic channel that we have found of application in calculating the particle current density, $J_\V{n}$, across the ESCs of an HCSC. We review this theory in more detail in the Appendix. 

According to this theory, the net particle current density across a mono-energetic channel can be regarded as the result of the addition of two ballistic particle current densities, one transporting particles with electrochemical potential $\mu_0$ and temperature $T_0$, with velocity $v$, from the absorber towards the terminal and, another one, transporting particles, characterized by the electrochemical potential $\mu_\V{T}$ and temperature $T_\V{T}$, and also at velocity $v$, from the terminal towards the absorber: 
\begin{eqnarray}\label{eq:jnideal}
J_\V{n}=\mathit{\Gamma} \left ( f[\epsilon-\mu_0,T_0]-f[\epsilon-\mu_\V{T},T_\V{T}] \right)      
\end{eqnarray}
where,
\begin{eqnarray}\label{eq:gammavelo1}
\mathit{\Gamma}=d_\epsilon v,
\end{eqnarray}
that we shall call \emph {saturation particle current density}, has units of m$^{-2}$s$^{-1}$. In Eq. \ref{eq:gammavelo1}, $d_\epsilon$ is the number of states per unit of volume available  to particles with energy $\epsilon$. For simplicty, $\mathit{\Gamma}$ will be assumed to be equal for both contacts. In Eq. \ref{eq:jnideal}, $f[\epsilon-\mu,T]$ is the Fermi-Dirac distribution probability of an electron which occupies a state with energy $\epsilon$ in a thermodynamic system characterized by the electrochemical potential $\mu$ and temperature $T$ (Eq. \ref{eq:Fermi}).
  
At this stage we observe that, when in section \ref{sec:ESC} we assumed ideal ESCs, Eqs. (\ref{eq:muH}) and (\ref{eq:muL}) had to be satisfied. These equations would lead us now to the right side of Eq. (\ref{eq:jnideal}) to be zero and, therefore, to a vanishing particle current density as Limpert and Bremner \cite{limpert_hot_2015} warned.
However, a non-vanishing particle current density can still be  admitted if  we accept that, in this limiting case, an ideal ESC demands an infinite number of states per unit of volume, $d_\epsilon$, or/and an infinite transport velocity, $v$. We shall later see that this ideal conditions also lead to ESCs  characterized by electric and thermal conductivity approaching infinity.

 %A non-vanishing particle current density for the isentropic transport %conditions anticipated by Eqs. \ref{eq:muH} or \ref{eq:muL} would require %$\mathit{\Gamma}\rightarrow\infty$ which leads us to the conclusion %(\ref{eq:gammavelo1}) that an ideal ESCs demands an infinite number of %states per unit of volume, $d_\epsilon$, or/and an infinite transport %velocity, $v$. 
 
 Eq. \ref{eq:jnideal} now allows us to calculate the current density vs voltage characteristic of an HCSC with  non-ideal ESCs,  that is, for $\mathit{\Gamma \neq \infty}$. For that, once $\Delta \epsilon$ and $\epsilon_\V{H}$ are fixed, for a given $T_0$, we can use,   first, Eqs. \ref{eq:Pequals} to \ref{eq:P_ou} to calculate $J_\V{n}$. Then, since this current density exits through the H-ESC, we can use:
 \begin{eqnarray}
 J_\V{n}=\mathit{\Gamma} \left ( f[\epsilon_\V{H}-\mu_0,T_0]-f[\epsilon_\V{H}-\mu_\V{H},T_\V{T}] \right)    
 \end{eqnarray}
 to calculate $\mu_\V{H}$ and, since this current has to return to the absorber through the L-ESC, and use:
  \begin{eqnarray}
 J_\V{n}=-\mathit{\Gamma} \left ( f[\epsilon_\V{L}-\mu_0,T_0]-f[\epsilon_\V{L}-\mu_\V{L},T_\V{T}] \right)    
 \end{eqnarray}
to calculate $\mu_\V{L}$. Once $\mu_\V{H}$ and $\mu_\V{L}$ have been calculated, Eq. \ref{eq:V} can be used to determine the output voltage of the HCSC and, together with $J_\V{n}$, the characteristic of the cell.
 
\begin{figure}[htp] 
\subfloat{%
  \includegraphics[clip,width=\columnwidth]{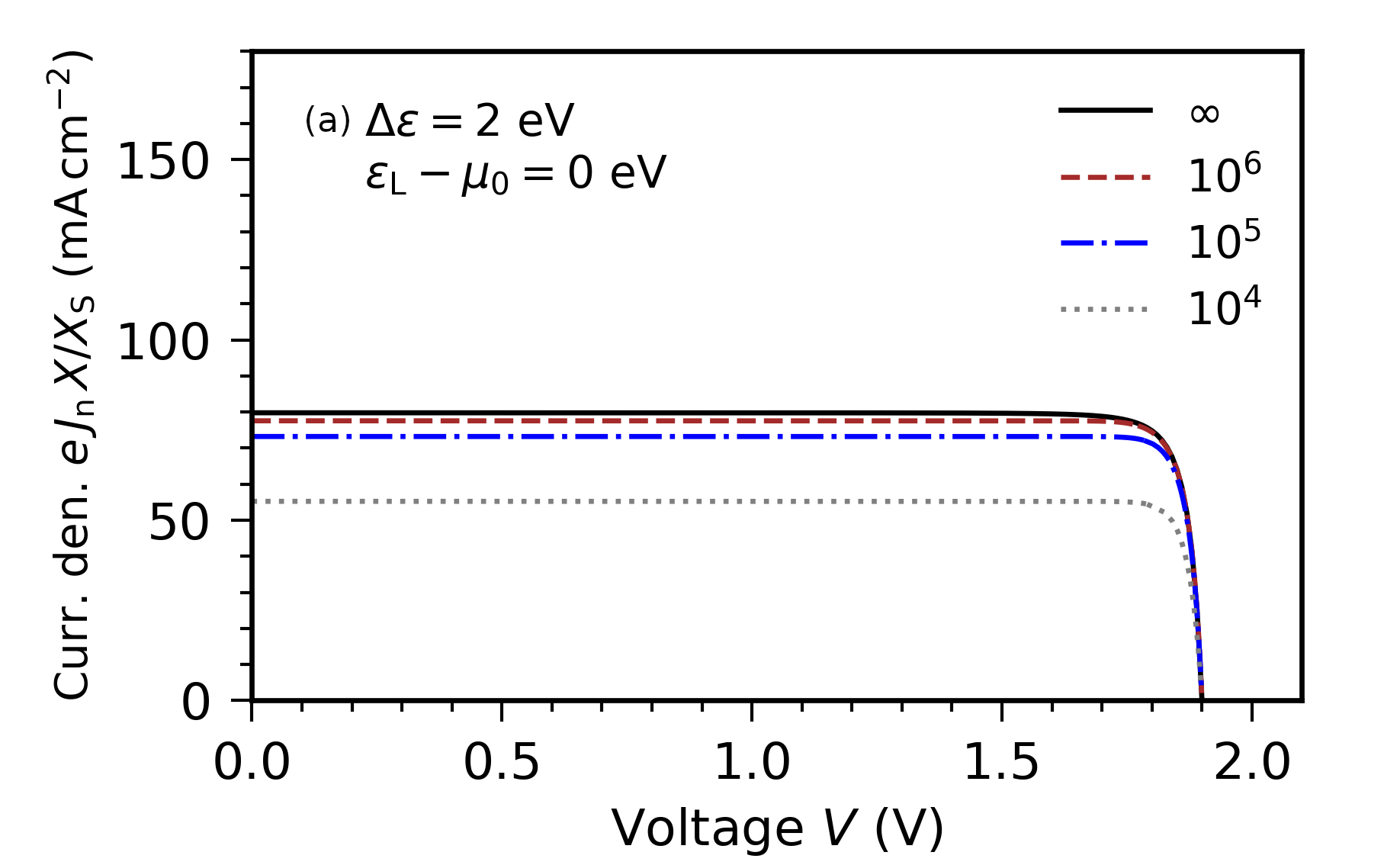}%
}

\subfloat{%
  \includegraphics[clip,width=\columnwidth]{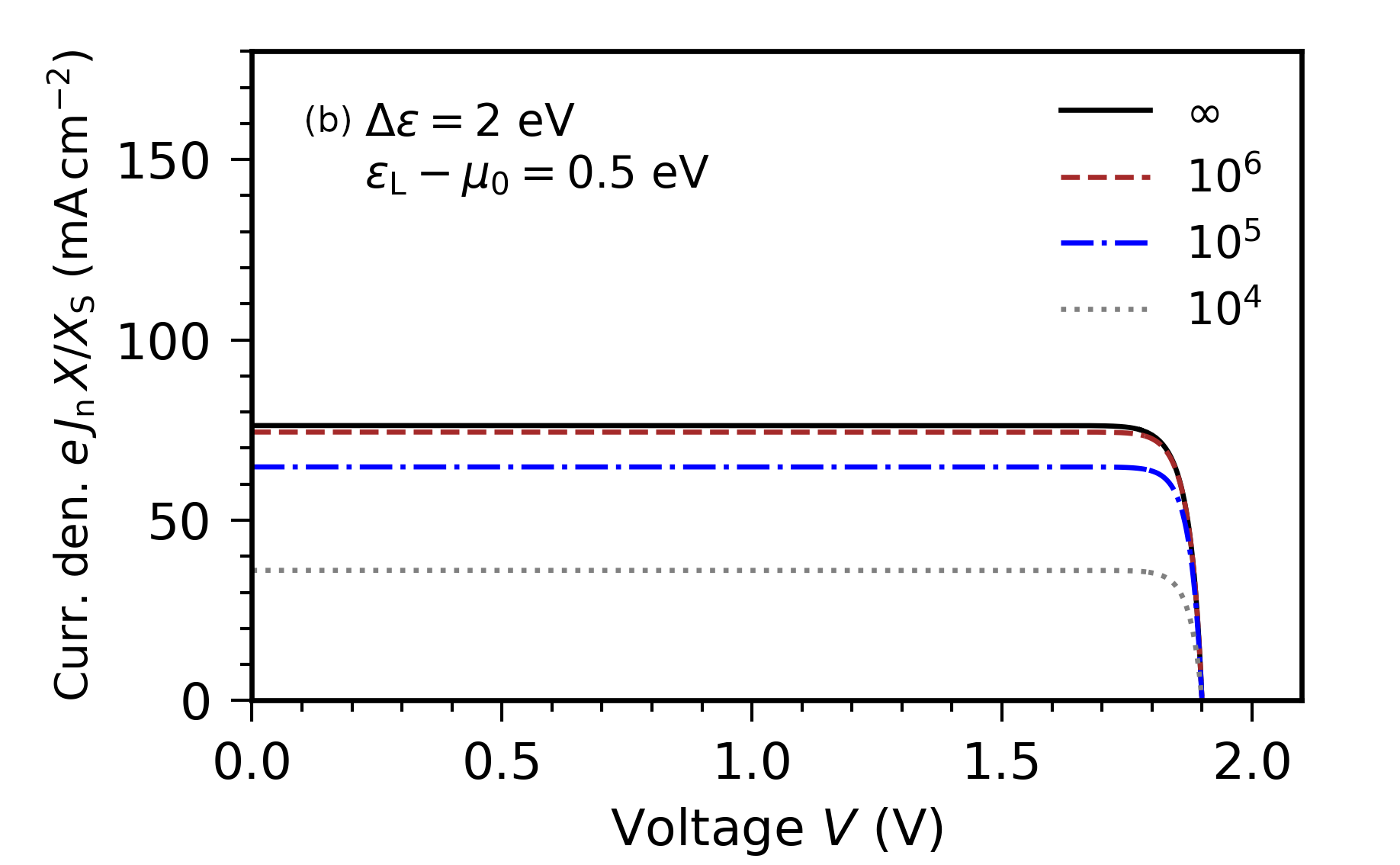}%   
}

\subfloat{%
  \includegraphics[clip,width=\columnwidth]{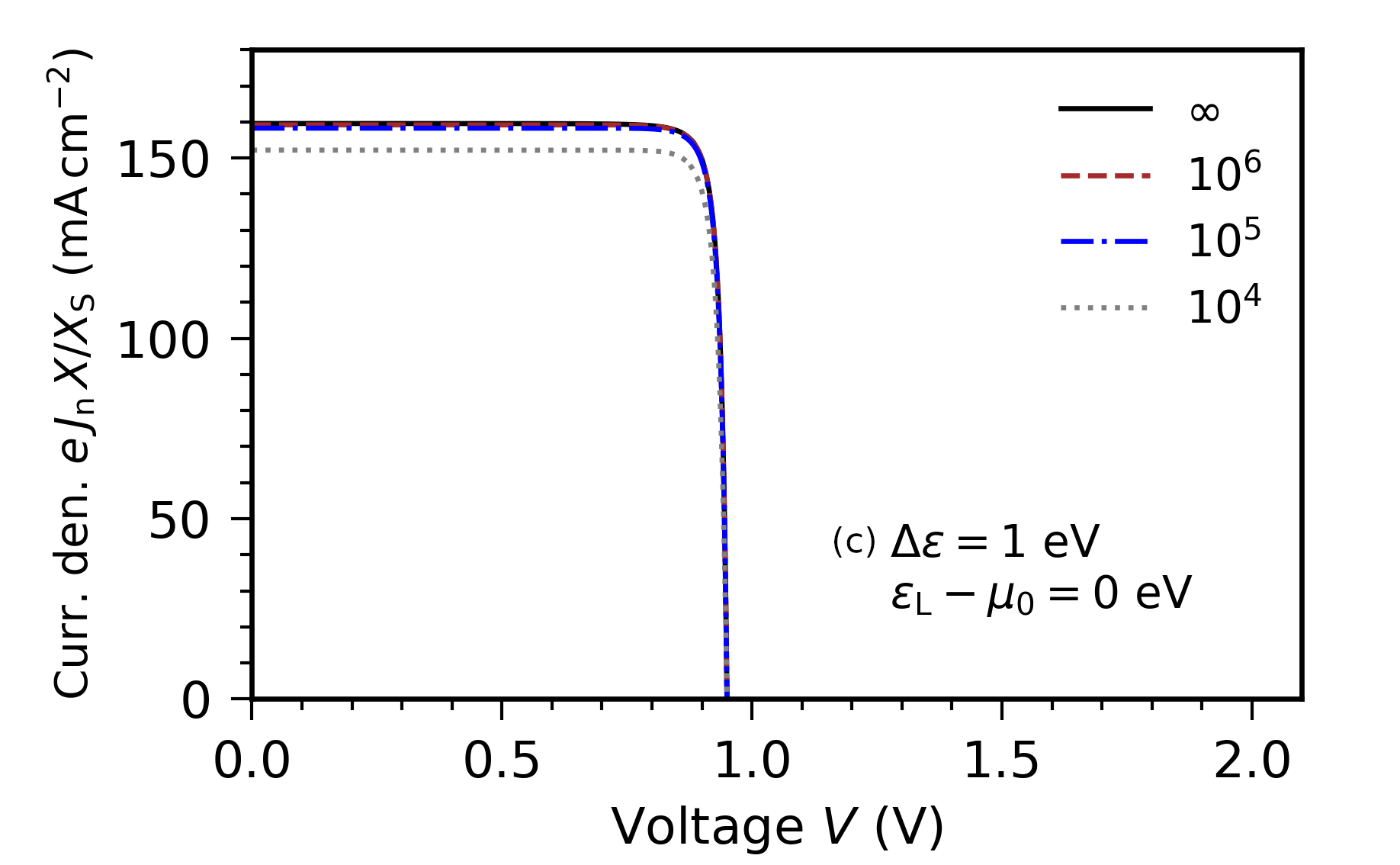}%
}

\caption{\label{fig:JV_gamma} Electric current density vs. voltage characteristics of a HCSC with non-ideal contacts for the following cases: (a) $\Delta \epsilon =2$ eV; $\epsilon_\V{L}-\mu_0=0$; (b) $\Delta \epsilon =2$ eV; $\epsilon_\V{L}-\mu_0=0.5$ eV; (c) $\Delta \epsilon =1$ eV; $\epsilon_\V{L}-\mu_0=0$ eV. The legend corresponds to the value of $e\mathit{\Gamma}/X_\V{S}$ in mA\kern 0.15emcm$^{-2}$.}

\end{figure}

Some examples of current density vs voltage characteristics thus  calculated have been plotted in (Fig. \ref{fig:JV_gamma}) and the corresponding efficiency calculated (Table \ref{tab:table_temp}). The examples include the case for which $\Delta \epsilon= 2$ eV and  $\epsilon_\V{L}-\mu_0=0$ eV  (Fig. \ref{fig:JV_gamma},a), the case for which $\Delta \epsilon= 2$ eV and  $\epsilon_\V{L}-\mu_0=0.5$ eV   (Fig. \ref{fig:JV_gamma},b) and the case  $\Delta \epsilon= 1$ eV and  $\epsilon_\V{L}-\mu_0=0$ eV  (Fig. \ref{fig:JV_gamma},c).

 % former JV_gamma,JV_4,JV_5
 
\begin{table}
\caption{\label{tab:table_temp} Temperatures at which short-circuit ($T_\V{SC}$), maximum power point ($T_\V{MX}$) and open-circuit ($T_\V{OC}$) conditions are attained in an HCSC with non-ideal energy selective contacts for several values of $\mathit{\Gamma}$ ($T_\V{S}=6000$ K; $T_\V{T}=300$ K). } 
\begin{ruledtabular}
\begin{tabular}{ccccc}
$e\mathit{\Gamma}/X_\V{S}$ [mA\kern 0.15emcm$^{-2}$] & $T_\V{SC}$ [K] & $T_\V{MX}$ [K] & $T_\V{OC}$ [K] & $\eta$\\
\hline
\multicolumn{5}{c}{ideal HCSC}\\
$\infty$ & 300 & 2544 & 6000 & 0.8536\\
\hline
\multicolumn{5}{c}{$\Delta \epsilon=2$ eV, $\epsilon_\V{L}-\mu_0=0$ eV}\\
10$^6$   & 2452 & 2760 & 6000 & 0.8462 \\  
10$^5$   & 3215 & 3340 & 6000 & 0.8065 \\
10$^4$   & 4469 & 4504 & 6000 & 0.6109\\
\hline 
\multicolumn{5}{c}{$\Delta \epsilon=2$ eV, $\epsilon_\V{L}-\mu_0=0.5$ eV }\\
10$^6$   & 3052 & 3195 & 6000 & 0.8210 \\ 
10$^5$   & 3951 & 4010 & 6000 & 0.7181 \\
10$^4$   & 5161 & 5176 & 6000 & 0.3999 \\
\hline
\multicolumn{5}{c}{$\Delta \epsilon=1$ eV, $\epsilon_\V{L}-\mu_0=0$ eV }\\
10$^6$   & 1327 & 2552 & 6000 & 0.8532 \\
10$^5$   & 1800 & 2617 & 6000 & 0.8496 \\
10$^4$   & 2783 & 3110 & 6000 & 0.8125 \\   
\end{tabular}
\end{ruledtabular}
\end{table}
 As expected, the current density plots in Fig. \ref{fig:JV_gamma}  reproduce Würfel's results represented in Fig. \ref{fig:JV} when $\mathit{\Gamma} \rightarrow \infty$. However, the efficiency of the HCSC decreases as $\mathit{\Gamma}$ decreases in agreement with the fact that finite values for $\mathit{\Gamma}$ correspond to non-isentropic transport through the ESCs. This is also in agreement with the results obtained by Limpert and Bremmer \cite{limpert_hot_2015} who previously pointed out a decrease in the HCSC efficiency when the isentropic conditions defined by Eqs. \ref{eq:muH} and \ref{eq:muL} were not fulfilled. Moreover, for the cases illustrated, we see (Table \ref{tab:table_temp}) that $e \mathit{\Gamma}/ X_\V{S}\ge 10^6$ mA\kern 0.15emcm$^{-2}$ is required to preserve the efficiency of an HCSC above 82 \%. Assuming $v=10^7$ cm\kern 0.15ems$^{-1}$ this would demand a material for the ESC capable of providing $d_\epsilon \approx 3 \times 10^{19}$ cm$^{-3}$ states available to the electrons.

The reduction of $\mathit{\Gamma}$ also impacts the short-circuit current of the cell but has no impact on its open-circuit voltage. In this respect, the impact of a decreasing $\mathit{\Gamma}$ would be similar to the increase in the series resistance of a solar cell. 
 
Decreasing $\mathit{\Gamma}$ also increases the temperature at which the maximum efficiency is obtained (Table \ref{tab:table_temp}). This result has implications for a practical HCSC, since a non-ideal HCSC, characterized, for example, by a non-ideal factor $e \mathit{\Gamma}/X_\V{S}=10^5$ mA\kern 0.15emcm$^{-2}$  would demand operation temperatures as high as $T_0=4000$ K, in the case $\Delta \epsilon=2$ eV and $\epsilon_\V{L}-\mu_0=0.5$ eV, in order to preserve an efficiency above 70\%.
 
We also observe that, while the limiting efficiency of an ideal HCSC did not depend on $\Delta \epsilon$, the limiting efficiency of an HCSC with non-ideal energy selective contacts depends, not only on $\Delta \epsilon$, but also on $\epsilon_\V{L}-\mu_0$. In this respect, we find out that the limiting efficiency of the HCSC decreases as the energy of the L-ESC, $\epsilon_\V{L}$, is located increasingly above $\mu_0$. The physical reason behind this result is that, as the energy of a non-ideal L-ESC increases with respect to $\mu_0$, it becomes increasingly difficult to populate its states with electrons and, therefore, to make it capable of transporting electric current. From these results we conclude that it is easier, in the sense of making an HCSC more tolerant to low values of $\mathit{\Gamma}$, to approach an ideal HCSC by reducing $\Delta \epsilon$ and making $\epsilon_\V{L}-\mu_0$ to approach zero.

\section{\label{sec:level1} Electric, thermal Conductivity, Seebeck coefficient and figure of merit of non-ideal energy selective contacts\protect }
As detailed in the Appendix, the theory based on the work by Datta and Landauer to calculate the particle current density of a non-ideal ESC allows us also to calculate the net heat current density, $J_\V{q}$,  and, from this knowledge together with $J_\V{n}$, also to calculate the electric conductivity $\sigma$ (Eq. \ref{eq:sigmaA}), thermal conductivity $\kappa$ (Eq. \ref{eq:kappaA}), Seebeck coefficient $\xi$ (Eq. \ref{eq:xiA}) and figure of merit $ZT$ (Eq. \ref{eq:ZTA}) of the ESC. These values depend not only on the energy of the channel of the ESC, but also on electrochemical potential and temperature. Therefore, for a given HCSC, we have different values for these parameters depending, first, on whether we are considering the hot side of the ESC (the absorber-channel side) or the cold side (the channel-terminal interface)  and, second, on whether we are considering the H-ESC or the L-ESC. We shall explore  the values these parameters take in the  following paragraphs.

\begin{figure*}
    \centering
    \hspace{0.22cm}
    \subfloat{%
        \includegraphics[clip,width=\columnwidth]{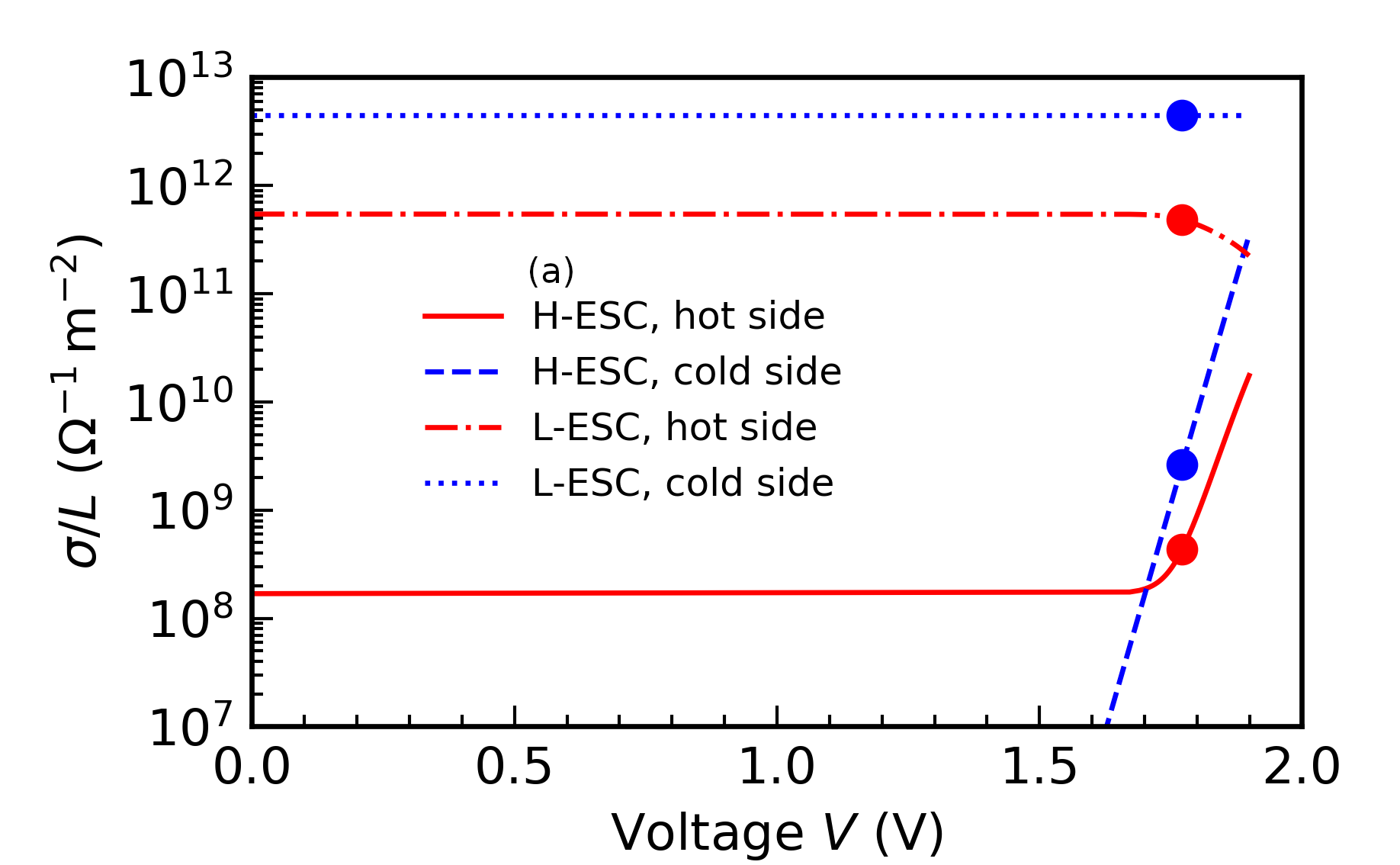}%
        }
    %\hfill
    \hspace{-0.1cm}
    \subfloat[]{%
        \includegraphics[clip,width=\columnwidth]{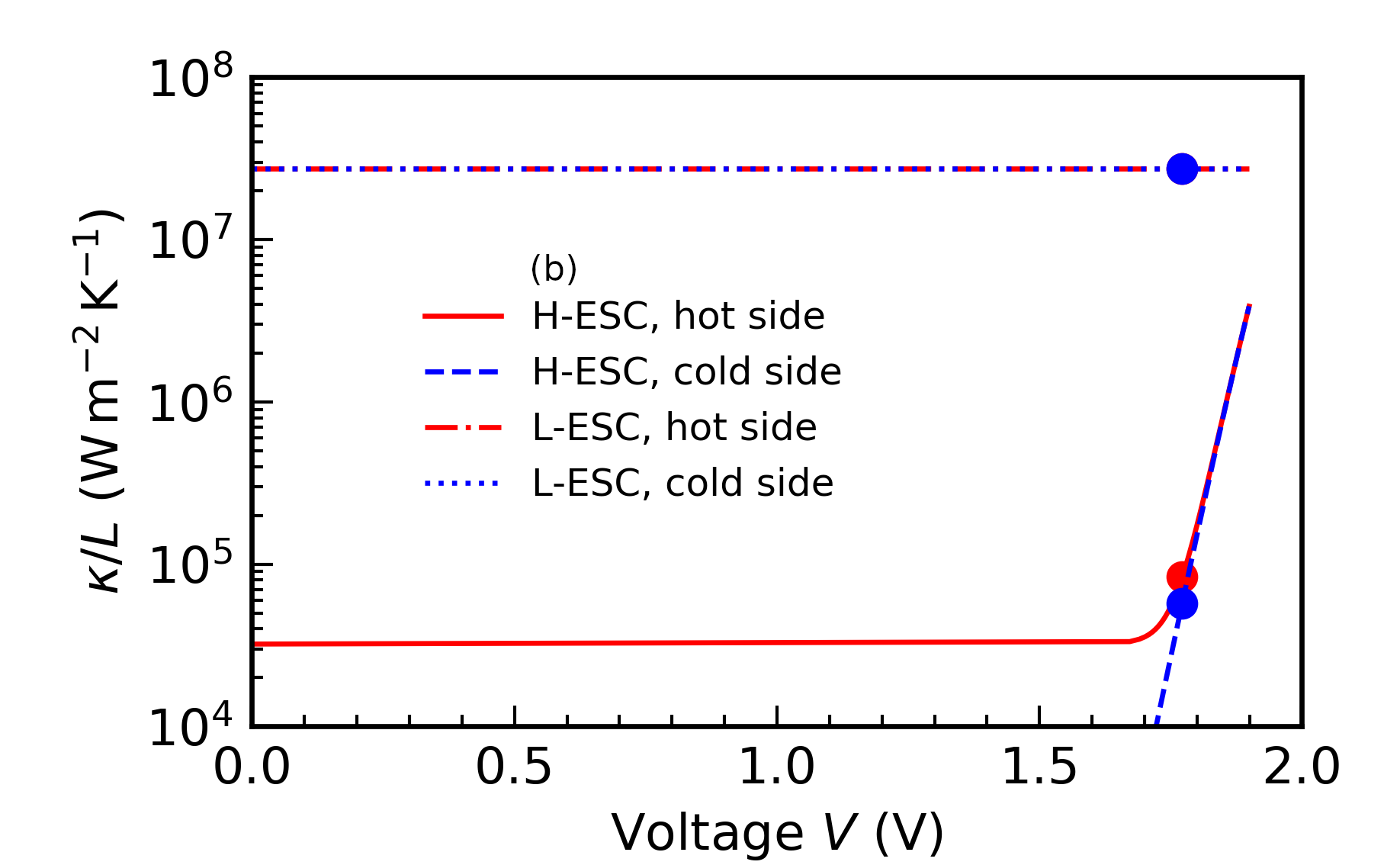}%
        }
    \hfill
    \vspace{0.2cm}
    \subfloat{%
        \includegraphics[clip,width=\columnwidth]{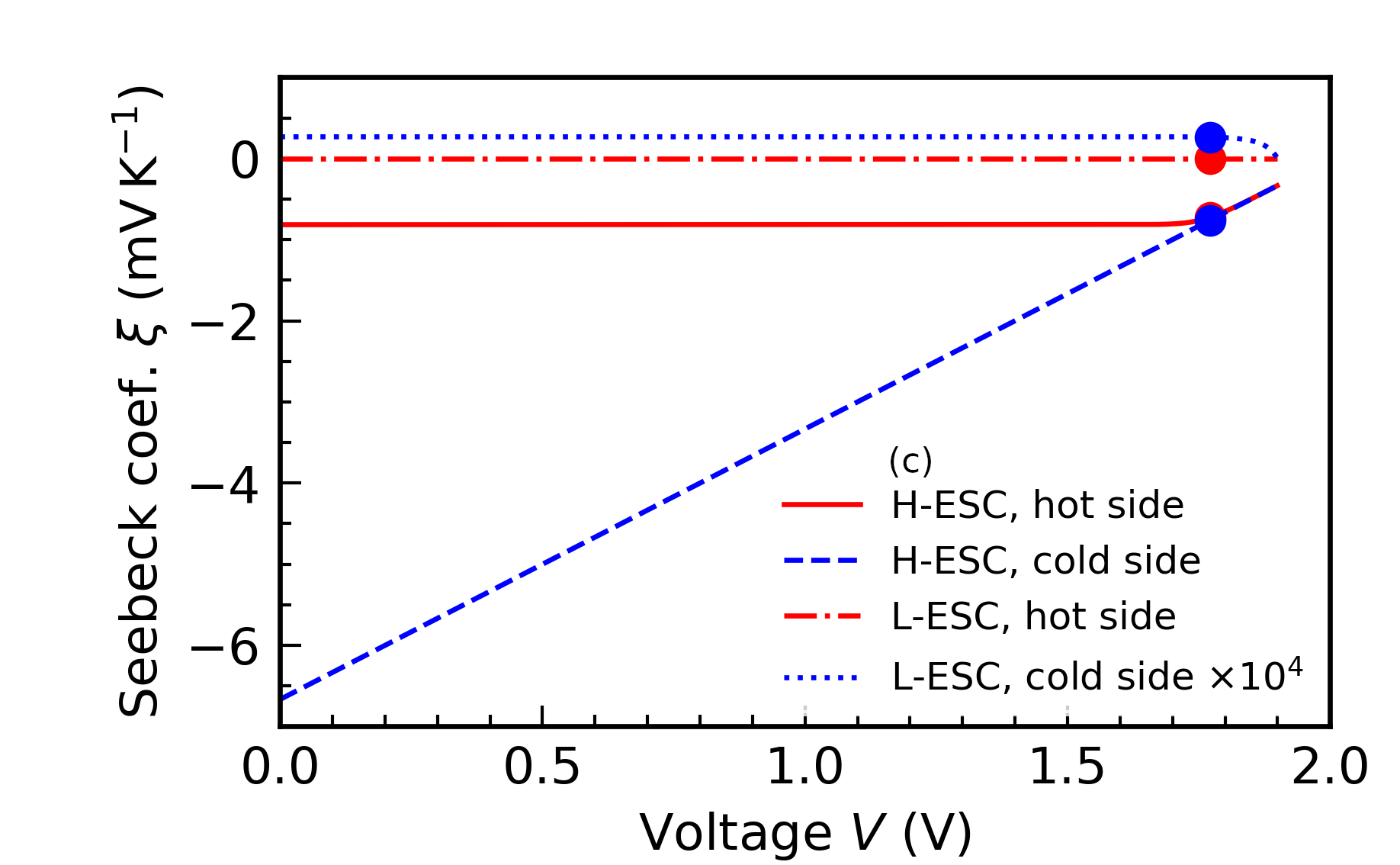}%
        }
    \hspace{-0.05cm}
    \subfloat{%
        \includegraphics[clip,width=\columnwidth]{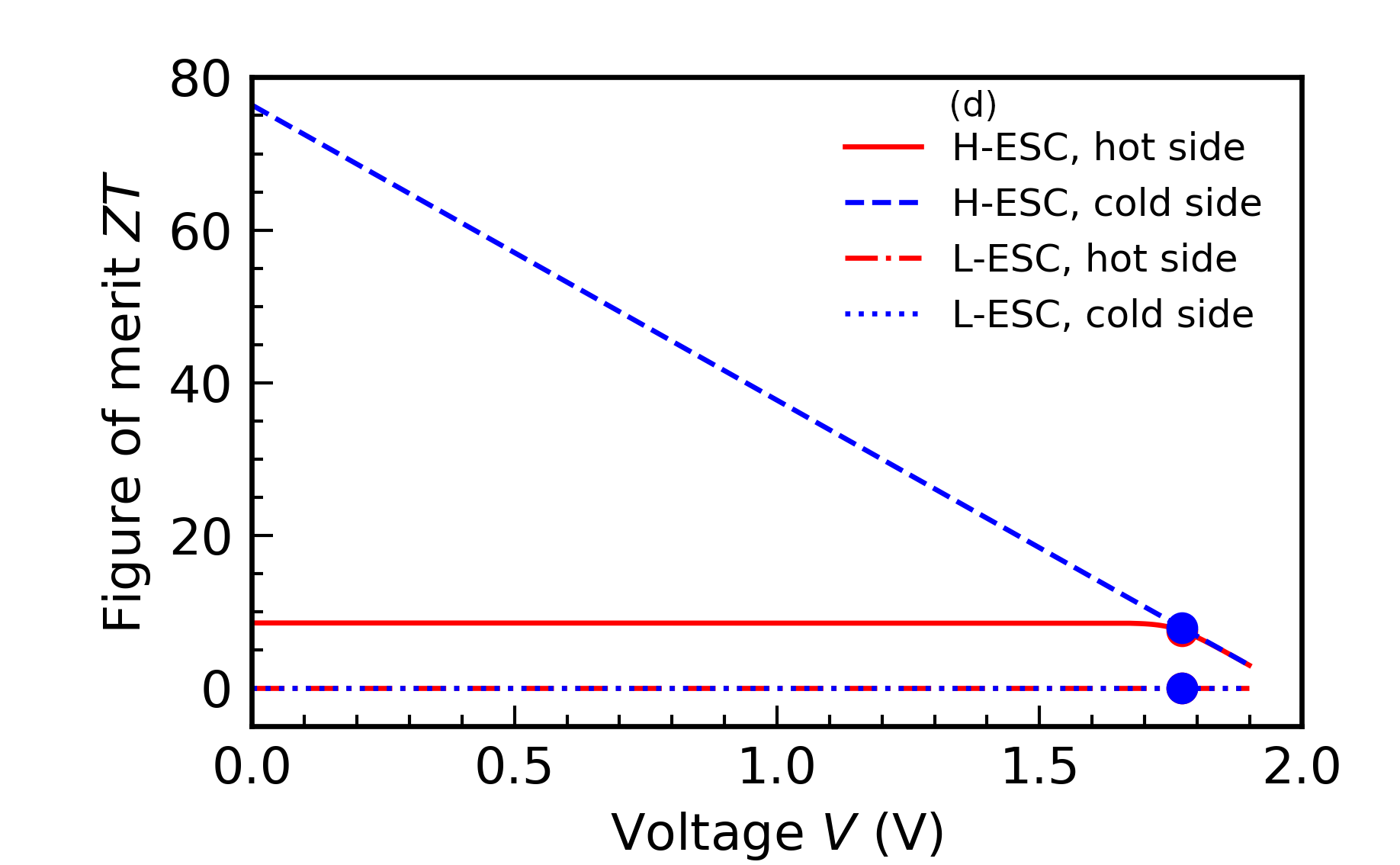}%
        }
    \caption{(a) Electric conductivity per unit of length, (b) thermal conductivity per unit of length, (c) absolute Seebeck coefficient and (d) figure of merit $ZT$ of the high energy (H-ESC) and low energy (L-ESC) selective contacts of a hot carrier solar cell as a function of the output voltage of the cell. The values are also plotted for the absorber-channel (hot side) and channel-terminal (cold side) of the contact. The contacts in this example are characterized by  $\Delta \epsilon=2$ eV,  $\epsilon_\V{L}-\mu_0=0$ and $e\mathit{\Gamma}/X_\V{S}=10^6$ mA\kern 0.15emcm$^{-2}$. The filled dots indicate the operation point at which the maximum electric power is extracted from the cell.}
    \label{fig:parameters}
\end{figure*}

Starting with the electric conductivity,  Fig. \ref{fig:parameters}a plots its value, as a function of the output voltage of the cell, for the case of an HCSC characterized by $e\mathit{\Gamma}/X_\V{S}=10^6$\kern 0.15emmA\kern0.15emcm$^{-2}$, $\Delta \epsilon=2$ 
eV and $\epsilon_\V{L}-\mu_0=0$ eV. We have chosen these values as representative ones because, in spite of the finite value of $\mathit{\Gamma}$, they preserve the limiting efficiency of the HCSC above 84 \% and a temperature of operation at the maximum power point below 3000 K. According to these results, and for solar cell operation at the maximum power point (indicated by a filled dot in the plots in the figure) we see that the most demanding contact in terms of high conductivity would be the cold side of the L-ESC  that requires $\sigma/L=4.5\times10^{12}$ $\Omega^{-1}$\kern 0.15emm$^{-2}$. We recall at this point (see the Appendix) that under the assumption of ballistic transport and mono-energetic channel, the electric conductivity depends on the length of the contact $L$ because the number of states available to electrons with energy $\epsilon$ increases linearly with $L$. In order to assist to the interpretation of the results we note that, if we were to compare it with silver, which has a conductivity of $\sigma=62.6\times 10^6$ $\Omega^{-1}$\kern 0.15emm$^{-1}$ at 295 K \citep{Landolt_r}, this result would tell us that the silver contact should be around 14 $\mu$m long in order to exhibit the required conductivity per unit of length.

%figure* utiliza toda la página

 Similarly, we plot in Fig. \ref{fig:parameters}b the thermal conductivity (which also depends on the length of the channel $L$) at the hot and cold sides of the H-ESC and L-ESC as a function of the output voltage for the same example of HCSC studied for the case of the electric conductivity.  

From these results we observe, for example, that with $\kappa/L=57\times 10^3$ W\kern 0.15emm$^{-2}$\kern 0.15emK$^{-1}$, the cold side of the H-ESC is the one demanding the lowest conductivity. If we take again for reference the case of silver,  with a thermal conductivity of 434 W\kern 0.15emm$^{-1}$K$^{-1}$ at 300 K \citep{Landolt_r}, this result would imply that we would need around 8 mm of silver to exhibit this thermal conductivity per unit of length.

Fig. \ref{fig:parameters}c now plots the Seebeck coefficient (or thermoelectric power) at the hot and cold sides of the L-ESC and H-ESC for the same solar cell. As discussed in the Appendix, this value only depends on the energy, electrochemical potential and temperature of the contact and, therefore, we consider it is the most representative value of the fundamental  properties of an ESC since, contrary to the conductivities, its value cannot be tailored by tuning the density of states nor the length of the channel. It must be taken into account though, that since the optimum temperature of operation of the absorber (see Table \ref{tab:table_temp}) depends on $\mathit{\Gamma}$, this dependence also impacts the value of $\xi$. Eq. \ref{eq:xiA} reveals that $\xi<0$ (conduction by electrons) when $\epsilon>\mu$ which, similarly to the conduction band of a semiconductor, is in agreement with an occupation probability of the level (Eq. \ref{eq:Fermi}) lower that 0.5 that indicates that most of the states of the ESC are empty. On the other hand, Eq. \ref{eq:xiA} also reveals that $\xi>0$ (conduction by holes) when $\epsilon<\mu$ which, similarly to the valence band of a semiconductor, is in agreement with an occupation probability of the level (Eq. \ref{eq:Fermi}) greater that 0.5 that indicates that most of the states of the ESC are occupied.

The most demanding contact of the HCSC in this respect is the high energy selective contact that demands (at the maximum power point)  760 and 724 $\mathrm{\mu}$V\kern 0.15emK$^{-1}$ for the cold (300 K) and hot (2760 K) side respectively (in absolute value). For comparison, giant Seebeck factors (850 $\mathrm{\mu}$V\kern 0.15emK$^{-1}$) have been reported, for example, for a two-dimensional electron gas in SrTiO$_3$ \citep{Giant_2007} for temperatures not exceeding 300 K.  Around 900 $\mathrm{\mu}$V\kern 0.15emK$^{-1}$ have been reported for SiB$_{14}$ \cite{Goto} at 1000 K. We are not aware of giant Seebeck factors being reported at 2500 K, the temperature at the hot side of the H-ESC, which illustrates the demanding thermoelectric properties of the ESCs. With respect to the L-ESC, we observe that $\xi=0$ at the hot side, which will be later justified (Eq. \ref{eq:xiA}) because in this example, at this side of the contact, $\epsilon_\V{L}=\mu_0$ and $\xi$ is slightly positive at the cold side because at this side,  $\mu_\V{L} \gtrsim \epsilon_\V{L}$.

Once the electric and thermal conductivities and the thermoelectric power have been calculated, the $ZT$ figure of merit (Eq. \ref{eq:ZTA}) illustrates their combined value that is required. $ZT$ does not depend explicitly on $\mathit{\Gamma}$ and therefore can also be considered a fundamental property of the ESC. The value of  $ZT$ required for a HCSC characterized by $\Delta \epsilon=2$ eV and $\epsilon_\V{L}-\mu_0=0$ is plotted in  Fig. \ref{fig:parameters}d for illustrative purposes.

The most demanding contact is again the H-ESC that requires $ZT \approx 8$ for both its cold side and hot side. Considering the difficulty in finding materials with ZT$>$2 \cite{zhao_panoscopic_2014} this requirement seems difficult. Conversely, developing mono-energetic contacts for ESCs could provide the insight for developing materials with a high $\xi$ as previously suggested by Mahan and Sofo \cite{mahan}. These authors reached this conclusion when investigating the optimum distribution of energy states leading to the best thermoelectric material. Later, Jeong et al. \cite{jeong_best_2012} have reviewed the best band structure for thermoelectric performance and included the impact of parameters such as the bandwidth and scattering in their analysis. Dehkordi et al.  \cite{mehdizadeh_dehkordi_thermoelectric_2015} extended the review to approaches that have been proposed and implemented experimentally such as the use of quantum confinement,  energy filtering,  resonant levels, etc. Nakpathomkun et al. \cite{nak_2010} have theoretically compared the performance of quantum dot systems, 1D ballistic conductors and thermionic emitters and concluded that the 1D ballistic conductor offered the highest achievable efficiency.

\section{Conclusions}

We have used the Datta-Landauer theory for mono-energetic contacts to model the performance of an HCSC with non-ideal energy selective contacts. This model assumes that electrons travel ballistically from the absorber to the terminal and do not lose their energy during transport (this latest assumption is shared by the ideal HCSC model). 

The non-ideality of the ESC can be accounted for by means of the particle saturation current density $\mathit{\Gamma}$ so that when $\mathit{\Gamma} \rightarrow \infty$, the ideal HCSC model is recovered. As expected, as $\mathit{\Gamma}$ decreases, the limiting efficiency of the HCSC decreases but, in addition, it has also been found that the optimum operation temperature of the hot carriers in the absorber increases. For example, it has been seen that for an HCSC with ESCs characterized by $\Delta \epsilon=2$ eV and  $\epsilon_\V{L}-\mu_0=0$, preserving a limiting efficiency above 84 \% demands $e \mathit{\Gamma}/46050 \gtrsim 10^6$ mA\kern 0.15emcm$^{-2}$ which also allows the hot carrier temperature of operation to be remain below 3000 K. Assuming $v=10^7$ cm\kern 0.15ems$^{-1}$ this would demand a material for the ESC capable of providing $d_\epsilon \approx 3 \times 10^{19}$ cm$^{-3}$ states available for the electrons. More demanding values for $\mathit{\Gamma}$ to preserve the HCSC  efficiency are required if $\Delta\epsilon$ is decreased.

The model has allowed us also to calculate the electric conductivity, thermal conductivity, Seebeck coefficient and figure of merit $ZT$ of the ESC as a function of the electron temperature and electrochemical potential. Since each ESC contact is in equilibrium at different temperatures and electrochemical potentials, depending on whether we consider the hot side (the absorber-channel interface) or the cold side (the channel-terminal interface) of the contact, different values for these magnitudes are obtained for each ESC and at each side of the contact. When the value of all these parameters is combined through the figure of merit $ZT$, the H-ESC becomes the most demanding when, for example, for $\Delta \epsilon= 2$ eV requires  $ZT \approx 8$ for both its cold side and hot side. We are not aware of any material exhibiting this value for $ZT$ which illustrates the difficulty for achieving a practical HCSC. On the other hand, the theory developed here is aligned with  previous works  supporting the idea that low dimensional structures, as the mono-energetic ESCs discussed here, should be pursued to obtain materials with high $ZT$.

In addition, given the impact of the length $L$ on the volume of the contact in this model and, therefore, on the number of states available to electrons in the channel, the actual value of the electric and thermal conductivity can be tuned by acting on $L$. However, neither the Seebeck factor nor the figure of merit $ZT$ depend on $L$, being their values fundamentally determined by the energy of the ESC and the electron temperature and electrochemical potential at which they have to operate.

\section{Methods}

Equations and results in this work have been obtained with the assistance of Mathematica.

\section{Appendix}
The calculation of the coefficients $L_{i,j}$, in Eqs. \ref{eq:Jn} and \ref{eq:JQ}, from which the electric conductivity, thermal conductivity, Seebeck coefficient and figure of merit $ZT$ can be derived, demands the calculation of the particle current density, $J_\V{n}$, and the heat current density, $J_\V{q}$ as a function of the particle temperature and electrochemical potential.

For the calculation of the particle current density we shall follow the approach by Datta \citep{Datta} that is based on a previous model by Landauer \cite{landauer_1957,landauer_1992}. To do so, first we must decide whether to consider the length of the channel, $L$, \emph{long} or \emph{short}. In this respect, since we want to preserve the mono-energetic nature of the channel in order to approach the ideal HCSC model, we choose the channel to be \emph{long} because this minimizes the interaction between the states of the absorber and the states of the electrons in the channel and, therefore, it also minimizes the broadening of the energetic band of the electrons in the channel as S. Datta discusses in his work.

Second, since we do not wish electrons in the channel to interact with phonons, in order to allow electrons to preserve their energy during their transport,  we shall also admit electrons to travel ballistically thorough the channel which is also a hypothesis of S.Datta's model.

With these basic hypotheses, the particle current density, $J_\V{n,1}$, injected from the absorber into the channel, in the direction of the channel can be written as:
\begin{eqnarray}\label{eq:jn1rev}
J_{\V{n},1}=\frac{1}{A_\V{C}}\frac{\gamma_1}{\hbar}\left(D_\epsilon f[\epsilon-\mu_0,T_0] -N\right)
\end{eqnarray}
where $A_\V{C}$ is the area of the contact; $D_\epsilon$ is the number of states in the channel available to electrons with energy $\epsilon$; $N$ is the average number of electrons in the channel;
\begin{eqnarray} \label{eq:Fermi}
f[\epsilon-\mu,T]=\frac{1}{1+\displaystyle \exp \left(\frac{\epsilon-\mu}{k_\V{B} T}\right )},
\end{eqnarray}
with $k_\V{B}$ being the Boltzmann constant, is the probability of an electron occupying an energy state $\epsilon$ in a thermodynamic system characterized by the electrochemical potential $\mu$ and temperature $T$, and $\gamma_1/\hbar$, with dimensions of s$^{-1}$, is the rate at which an electron located in the channel would escape into the absorber. 

Similarly, the particle current density, $J_\V{n,2}$, injected from the terminal into the channel, in the direction of the channel,  can be written as:  
\begin{eqnarray}\label{eq:jn2rev}
J_{\V{n},2}=\frac{1}{A_\V{C}}\frac{\gamma_2}{\hbar}\left(D_\epsilon f[\epsilon-\mu_\V{T},T_\V{T}] -N\right)
\end{eqnarray}
where $\gamma_2/\hbar$ is the rate at which an electron located in the channel would escape into the terminal. Note that, when dealing with the H-ESC, we have $\mu_\V{T}=\mu_\V{H}$ and $\mu_\V{T}=\mu_\V{L}$ when considering the L-ESC. 

On the other hand, $D_\epsilon$ can also be written as:
\begin{eqnarray} \label{eq:densityest}
D_\epsilon=d_\epsilon \hat{V}
\end{eqnarray}
where $d_\epsilon$ is the density of states per unit of volume available to electrons in the channel and
\begin{eqnarray}
\hat{V}=A_\V{C} L
\end{eqnarray}
is the volume of the channel. In addition, when the channel is \emph{long} and transport is ballistic,
\begin{eqnarray}
\frac{\gamma_1}{\hbar}=\frac{v_{\V{AC}}}{L} \\
\frac{\gamma_2}{\hbar}=\frac{v_{\V{TC}}}{L}
\end{eqnarray}
where $v_{\V{AC}}$ and $v_{\V{TC}}$ are a \emph{kind of surface velocity} at the absorber-channel and channel-terminal interface respectively.

Since no electron is created nor destroyed during its transport through the channel, the particle current density must be constant and, therefore:
\begin{eqnarray}
J_{\V{n},1}=-J_{\V{n},2}
\end{eqnarray}
which allows $N$ to be elliminated from Eqs. \ref{eq:jn1rev} and \ref{eq:jn2rev} to obtain:
\begin{eqnarray}\label{eq:jnrev}
J_\V{n}=J_{\V{n},1}=-J_{\V{n},2}=\mathit{\Gamma} \left ( f[\epsilon-\mu_0,T_0]-f[\epsilon-\mu_\V{T},T_\V{T}] \right) \nonumber \\
\end{eqnarray}
with
\begin{eqnarray}\label{eq:gammavelo}
\mathit{\Gamma}=d_\epsilon\frac{ v_{\V{AC}} v_{\V{TC}}}{v_{\V{AC}}+v_{\V{TC}}}
\end{eqnarray}
Note that $\mathit{\Gamma}$, that we shall define as the \emph{saturation particle current density} of the energy selective contact and that has units of m$^{-2}$s$^{-1}$, represents, since $f[\epsilon-\mu,T]\leq 1$, the maximum particle current density that the energy selective contact can sustain. 

Eq. \ref{eq:jnrev} can also be written as:
\begin{eqnarray}\label{eq:jnvelocity}
J_\V{n}=n[\epsilon-\mu_0,T_0] v-n[\epsilon-\mu_\V{T},T_\V{T}] v
\end{eqnarray}
with
\begin{eqnarray}
n[\epsilon-\mu_0,T_0]=&d_\epsilon f[\epsilon-\mu_0,T_0] \\
n[\epsilon-\mu_\V{T},T_\V{T}]=&d_\epsilon f[\epsilon-\mu_\V{T},T_\V{T}] \\
v=&\displaystyle\frac{v_{\V{AC}}v_{\V{AT}}}{v_{\V{AC}}+v_{\V{AT}}}
\end{eqnarray}
This allows us to physically interpret the particle current density thorugh the energy selective contact as the net result of a concentration of electrons $n[\epsilon-\mu_0,T_0]$ traveling ballistically from the absorber towards the terminal at speed $v$ \emph{minus} a concentration of electrons $n[\epsilon-\mu_\V{T},T_\V{T}]$ traveling ballistically back from the terminal towards the absorber also at speed $v$. In the following paragraphs we shall stick to this interpretation to calculate several thermodynamic quantities of interest and see how far it can take us in order to improve our understanding of the thermodynamic properties that the ESCs in an HCSC must exhibit.

An  experimentalist ignoring the details of the model might be willing to characterize these contacts under low excitation conditions and fit the results to the linear model  described by Eqs. \ref{eq:Jn} and \ref{eq:JQ}. In order to obtain the equivalent parameters $L_{1,1}$ and $L_{1,2}$, we first write the following trivial relationships:
\begin{eqnarray}
\mu_\V{T}=\mu_0+\frac{\mu_\V{T}-\mu_0}{L}L \label{eq:muT}\\
\frac{1}{T_\V{T}}=\frac{1}{T_\V{0}}+\frac{1/T_\V{T}-1/T_0}{L}L \label{eq:1oT}
\end{eqnarray}
and define
\begin{eqnarray}
\dot{\mu}=\frac{\mu_\V{T}-\mu_0}{L} \\
\dot{\mathcal{T}}=\frac{1/T_\V{T}-1/T_0}{L}
\end{eqnarray}
In a linear regimen, for which $\mu_\V{T}\approx\mu_0$ and $1/T_\V{T} \approx 1/T_0$, an observer would interpret, 
\begin{eqnarray}
\frac{d\mu}{dx}\approx \dot{\mu}\\
\frac{d(1/T)}{dx}\approx \dot{\mathcal{T}}
\end{eqnarray}

By inserting Eqs. \ref{eq:muT} and \ref{eq:1oT} into Eq. \ref{eq:jnrev} we obtain, for this linear regime in which $\dot{\mu}\rightarrow 0$ and $\dot{\mathcal{T}}\rightarrow0 $, that the particle current density, at the absorber-channel contact could be approximated by:
\begin{eqnarray}
J_\V{n} \approx -\frac{L_{1,1}[\epsilon-\mu_0,T_0]}{T_0}   \dot{\mu}-L_{1,2}[\epsilon-\mu_0,T_0]\dot{\mathcal{T}}
\end{eqnarray}
with
\begin{eqnarray}
L_{1,1}[\epsilon-\mu,T]=\frac{L \mathit{\Gamma}   \mathrm{sech} \left(\displaystyle \frac{\epsilon-\mu}{2 k_\V{B} T} \right)^2 }{4 k_\V{B}} \\
L_{1,2}[\epsilon-\mu,T]=-\frac{L \mathit{\Gamma} (\epsilon-\mu)   \mathrm{sech} \left(\displaystyle \frac{\epsilon-\mu}{2 k_\V{B} T} \right)^2 }{4 k_\V{B}}
\end{eqnarray}
Following a similar procedure, in which we insert:
\begin{eqnarray}
\mu_0=\mu_\V{T}-\frac{\mu_\V{T}-\mu_0}{L}L \label{eq:mu0}\\
\frac{1}{T_0}=\frac{1}{T_\V{T}}-\frac{1/T_\V{T}-1/T_0}{L}L \label{eq:1oT0}
\end{eqnarray}
into Eq. \ref{eq:jnrev}, we obtain that the particle current density at the channel-terminal interface could be approximated, in the linear regime, by:
\begin{eqnarray}
J_\V{n} \approx -\frac{L_{1,1}[\epsilon-\mu_\V{T},T_\V{T}]}{T_\V{T}}   \dot{\mu}-L_{1,2}[\epsilon-\mu_\V{T},T_\V{T}]\dot{\mathcal{T}} \nonumber \\
\end{eqnarray}

For the calculation of the coefficients $L_{2,1}$ and $L_{2,2}$ we need to calculate the heat current density, $J_\V{q}$. Following the ballistic approach, this current density is given by the heat transported by the particle concentration $n[\mu_0,T_0]$, that travels from the absorber towards the terminal minus the heat transported by the particle concentration $n[\mu_\V{T},T_\V{T}]$ traveling from the terminal towards the absorber:
\begin{eqnarray} \label{eq:jqentropy}
J_\V{q}=T_0 s[\epsilon-\mu_0,T_0] v-T_\V{T} s[\epsilon-\mu_\V{T},T_\V{T}] v
\end{eqnarray}

In this equation,  $s[\epsilon-\mu,T]$ is the entropy per unit of volume transported by the concentration of particles $n[\epsilon-\mu,T]$. This entropy is given by \citep{Luque_Handbook}:
\begin{eqnarray}
s[\epsilon-\mu,T]=\frac{1}{T} \hat{e}[\epsilon-\mu,T]-\frac{\mu}{T} n[\epsilon-\mu,T]+\frac{p[\epsilon-\mu,T]}{T} \nonumber \\
\end{eqnarray}
where $\hat{e}[\epsilon-\mu,T]$ is the energy per unit of volume transported by the particle concentration $n[\epsilon-\mu,T]$,
\begin{eqnarray}
\hat{e}[\epsilon-\mu_,T]=\epsilon n[\epsilon-\mu,T]
\end{eqnarray}
and  $p[\epsilon-\mu,T]$, their pressure. For the case of an electron gas, with only one energy  $\epsilon$ possible for each electron and with a degeneracy of $d_\epsilon$ states per unit of volume, this pressure is given by:
\begin{eqnarray}
p[\epsilon-\mu,T]=d_\epsilon k_\V{B} T \ln \left(1+\exp\frac{\mu-\epsilon}{k_\V{B} T}\right)
\end{eqnarray}

To obtain the coefficients $L_{2,1}$ and $L_{2,2}$ at the absorber-channel side, we insert Eqs. \ref{eq:muT}
and \ref{eq:1oT} into Eq. \ref{eq:jqentropy} and linearize for $\dot{\mu}\rightarrow 0$ and $\dot{\mathcal{T}}\rightarrow 0$. Hence, for this linear regimen we obtain that the heat current density at the absorber-channel side would be given approximately by:
\begin{eqnarray}
J_\V{q} \approx \frac{L_{2,1}[\epsilon-\mu_0,T_0]}{T_0} \dot{\mu}+L_{2,2}[\epsilon-\mu_0,T_0] \dot{\mathcal{T}} \nonumber \\
\end{eqnarray}
with
\begin{widetext}
\begin{eqnarray}
L_{2,1}[\epsilon-\mu,T]=&-\displaystyle \frac{L \mathit{\Gamma} (\epsilon-\mu)   \mathrm{sech} \left(\displaystyle \frac{\epsilon-\mu}{2 k_\V{B} T} \right)^2 }{4 k_\V{B}} \\
L_{2,2}[\epsilon-\mu,T]=&\mathit{\Gamma}L T 
\left(\displaystyle \frac{(\epsilon-\mu)\left(1+\left(1+\frac{\epsilon-\mu}{k_\V{B} T} \right)\exp \frac{\epsilon-\mu}{k_\V{B} T} \right)}{ \left(1+\exp \frac{\epsilon-\mu}{k_\V{B} T}\right)^2} + k_\V{B} T\ln \left(1+\exp \frac{\mu-\epsilon}{k_\V{B} T} \right) \right)
\end{eqnarray}
\end{widetext}

Proceeding similarly by inserting Eqs. \ref{eq:mu0} and \ref{eq:1oT0} into \ref{eq:jqentropy} and linearizing  we can see that, at the channel-terminal interface, the heat current density could be written as:
  \begin{eqnarray}
J_\V{q} \approx =\frac{L_{2,1}[\epsilon-\mu_\V{T},T_\V{T}]}{T_\V{T}} \dot{\mu}+L_{2,2}[\epsilon-\mu_\V{T},T_\V{T}] \dot{\mathcal{T}}
\end{eqnarray}

Note that we have obtained $L_{2,1}=L_{1,2}$ which satisfies the Onsager relationship for Markovian systems (systems that do not depend on their previous history) \citep{Callen2}
which provides additional support to the physical consistency of the model.

Once the coefficients $L_{i,j}$ have been obtained, the electric conductivity $\sigma$, thermal conductivity, $\kappa$, Seebeck coefficient $\xi$ and figure of merit $ZT$ can be calculated using the following relationships \citep{Callen1} (note that in this reference $e$ implies $+e$ for holes and $-e$ for electrons while in our work $e$ is always in absolute value):
\begin{widetext}
\begin{eqnarray}
\sigma[\epsilon-\mu,T]=& \displaystyle \frac{e^2 L_{1,1}}{T}=\displaystyle \frac{\Gamma L e^2 }{4 k_\V{B} T} \mathrm{sech} \left[\frac{(\epsilon-\mu)}{2 k_\V{B} T}\right]^2 \label{eq:sigmaA} \\
\kappa[\epsilon-\mu,T]=& \displaystyle \frac{L_{1,1}L_{2,2}-L_{1,2}^2}{T^2L_{1,1}} = 
\Gamma L \left(  \displaystyle \frac{\epsilon-\mu}{T} \frac{1}{1+\exp\frac{\epsilon-\mu}{k_\V{B} T}} +k_\V{B} \ln \left(1+ \displaystyle \frac{-\epsilon+\mu}{k_\V{B} T}   \right)  \right) \nonumber \\ \label{eq:kappaA} \\
\xi[\epsilon-\mu,T]=&\displaystyle  \frac{L_{1,2}}{eTL_{1,1}}=- \frac{\epsilon-\mu}{T e}  \label{eq:xiA} \\
ZT[\epsilon-\mu,T]=&\displaystyle \frac{\sigma \xi^2 T}{\kappa}=
\displaystyle \frac{(\epsilon-\mu)^2 \mathrm{sech}\left[ \displaystyle \frac{\epsilon-\mu}{2 k_\V{B} T}\right]^2}{4 k_\V{B} T\left(\displaystyle \frac{\epsilon -\mu}{1+\exp \displaystyle \frac{\epsilon -\mu}{k_\V{B} T}}+k_\V{B} T \ln \left[ 1+\exp \frac{-\epsilon+\mu}{k_\V{B} T}\right]\right)} \label{eq:ZTA}
\end{eqnarray}
\end{widetext}

Note that we shall have different values for these parameters, depending on whether we  consider the absorber-channel or the channel-terminal interface, as the electrochemical potential and temperature are different at each side of the channel and, furthermore, the values differ depending on whether we consider the H-ESC or the L-ESC, since the electron energy $\epsilon$ is different in each channel.

On the other hand,  the dependence of $\sigma$ and $\kappa$ on the geometric parameter $L$ might be surprising. As pointed out by S.Datta \citep{Datta}, this a consequence of the ballistic transport assumption and the fact that the total density of states $D_\epsilon$ available for the electrons depends on the length $L$  (Eq. \ref{eq:densityest}). The consequence of this is that $\sigma$ and $\kappa$ can be ideally tuned by modifying the length of the contact $L$ as long as this modification does not cause the ballistic transport assumption to break-down. On the contrary, the Seebeck coefficient and $ZT$ cannot be tuned and become independent, not only on any geometric property, but also  on the density of states. Their value is only determined by the energy of the electrons,  the temperature and the electrochemical potential of the particles in the channel, and therefore they can be considered a fundamental property that the ESCs of the HCSC must exhibit. 

As a final remark, we observe that the net entropy current density across the channel, $J_\V{s}$, is given by:
\begin{eqnarray}
J_\V{s}[\epsilon-\mu,T]=s[\epsilon-\mu_0,T_0]v-s[\epsilon-\mu_\V{T},T_\V{T}] v
\end{eqnarray}
To maximize the work produced by the converter, the entropy current density exiting the H-ESC must equal the entropy current density entering the L-ESC: 
\begin{eqnarray}
 s[\epsilon_\V{H}-\mu_0,T_0]v-s[\epsilon_\V{H}-\mu_\V{H},T_\V{T}] v= \nonumber \\ s[\epsilon_\V{L}-\mu_\V{L},T_0]v-s[\epsilon_\V{L}-\mu_0,T_0] v
\end{eqnarray}
 This equality is satisfied for:
 \begin{eqnarray}
\mu_\V{H}-\mu_\V{L}=(\epsilon_\V{H}-\epsilon_\V{L})\left(1-\frac{T_\V{T}}{T_0}\right) 
 \end{eqnarray}
 which, of course, leads to the same Carnot efficiency that we obtained in Eq. \ref{eq:etaB}.

\section{Acknowledgments}
This work has been sponsored by the Project CEOTRES-CM (Y2020/EMT-6419) funded by the Comunidad de Madrid. I. Ramiro acknowledges funding of the European Commission through a H2020 Marie Skłodowska-Curie Action (GA 891686). E.A. acknowledges grant RYC-2015-18539 (Ram\'on y Cajal Fellowship), funded by MCIN/AEI/10.13039/501100011033 and by “ESF Investing in your future”, and the funding by Fundaci\'on Ram\'on Areces within
the research project SuGaR.

\bibliography{HCSC_2021_08}% Produces the bibliography via BibTeX.

\begin{thebibliography}{31}%
\makeatletter
\providecommand \@ifxundefined [1]{%
 \@ifx{#1\undefined}
}%
\providecommand \@ifnum [1]{%
 \ifnum #1\expandafter \@firstoftwo
 \else \expandafter \@secondoftwo
 \fi
}%
\providecommand \@ifx [1]{%
 \ifx #1\expandafter \@firstoftwo
 \else \expandafter \@secondoftwo
 \fi
}%
\providecommand \natexlab [1]{#1}%
\providecommand \enquote  [1]{``#1''}%
\providecommand \bibnamefont  [1]{#1}%
\providecommand \bibfnamefont [1]{#1}%
\providecommand \citenamefont [1]{#1}%
\providecommand \href@noop [0]{\@secondoftwo}%
\providecommand \href [0]{\begingroup \@sanitize@url \@href}%
\providecommand \@href[1]{\@@startlink{#1}\@@href}%
\providecommand \@@href[1]{\endgroup#1\@@endlink}%
\providecommand \@sanitize@url [0]{\catcode `\\12\catcode `\$12\catcode
  `\&12\catcode `\#12\catcode `\^12\catcode `\_12\catcode `\%12\relax}%
\providecommand \@@startlink[1]{}%
\providecommand \@@endlink[0]{}%
\providecommand \url  [0]{\begingroup\@sanitize@url \@url }%
\providecommand \@url [1]{\endgroup\@href {#1}{\urlprefix }}%
\providecommand \urlprefix  [0]{URL }%
\providecommand \Eprint [0]{\href }%
\providecommand \doibase [0]{https://doi.org/}%
\providecommand \selectlanguage [0]{\@gobble}%
\providecommand \bibinfo  [0]{\@secondoftwo}%
\providecommand \bibfield  [0]{\@secondoftwo}%
\providecommand \translation [1]{[#1]}%
\providecommand \BibitemOpen [0]{}%
\providecommand \bibitemStop [0]{}%
\providecommand \bibitemNoStop [0]{.\EOS\space}%
\providecommand \EOS [0]{\spacefactor3000\relax}%
\providecommand \BibitemShut  [1]{\csname bibitem#1\endcsname}%
\let\auto@bib@innerbib\@empty
%</preamble>
\bibitem [{\citenamefont {Ross}\ and\ \citenamefont {Nozik}(1982)}]{RossNozik}%
  \BibitemOpen
  \bibfield  {author} {\bibinfo {author} {\bibfnamefont {R.~T.}\ \bibnamefont
  {Ross}}\ and\ \bibinfo {author} {\bibfnamefont {A.~J.}\ \bibnamefont
  {Nozik}},\ }\bibfield  {title} {\bibinfo {title} {Efficiency of hot‐carrier
  solar energy converters},\ }\href@noop {} {\bibfield  {journal} {\bibinfo
  {journal} {Journal of Applied Physics}\ }\textbf {\bibinfo {volume} {53}},\
  \bibinfo {pages} {3813} (\bibinfo {year} {1982})}\BibitemShut {NoStop}%
\bibitem [{\citenamefont {Würfel}(1997)}]{WURFEL199743}%
  \BibitemOpen
  \bibfield  {author} {\bibinfo {author} {\bibfnamefont {P.}~\bibnamefont
  {Würfel}},\ }\bibfield  {title} {\bibinfo {title} {Solar energy conversion
  with hot electrons from impact ionisation},\ }\href
  {https://doi.org/https://doi.org/10.1016/S0927-0248(96)00092-X} {\bibfield
  {journal} {\bibinfo  {journal} {Solar Energy Materials and Solar Cells}\
  }\textbf {\bibinfo {volume} {46}},\ \bibinfo {pages} {43 } (\bibinfo {year}
  {1997})}\BibitemShut {NoStop}%
\bibitem [{\citenamefont {Knig}\ \emph {et~al.}(2010)\citenamefont {Knig},
  \citenamefont {Casalenuovo}, \citenamefont {Takeda}, \citenamefont
  {Conibeer}, \citenamefont {Guillemoles}, \citenamefont {Patterson},
  \citenamefont {Huang},\ and\ \citenamefont {Green}}]{knig_hot_2010}%
  \BibitemOpen
  \bibfield  {author} {\bibinfo {author} {\bibfnamefont {D.}~\bibnamefont
  {Knig}}, \bibinfo {author} {\bibfnamefont {K.}~\bibnamefont {Casalenuovo}},
  \bibinfo {author} {\bibfnamefont {Y.}~\bibnamefont {Takeda}}, \bibinfo
  {author} {\bibfnamefont {G.}~\bibnamefont {Conibeer}}, \bibinfo {author}
  {\bibfnamefont {J.}~\bibnamefont {Guillemoles}}, \bibinfo {author}
  {\bibfnamefont {R.}~\bibnamefont {Patterson}}, \bibinfo {author}
  {\bibfnamefont {L.}~\bibnamefont {Huang}},\ and\ \bibinfo {author}
  {\bibfnamefont {M.}~\bibnamefont {Green}},\ }\bibfield  {title} {\bibinfo
  {title} {Hot carrier solar cells: {Principles}, materials and design},\
  }\href {https://doi.org/10.1016/j.physe.2009.12.032} {\bibfield  {journal}
  {\bibinfo  {journal} {Physica E: Low-Dimensional Systems and Nanostructures}\
  }\textbf {\bibinfo {volume} {42}},\ \bibinfo {pages} {2862} (\bibinfo {year}
  {2010})}\BibitemShut {NoStop}%
\bibitem [{\citenamefont {König}\ \emph {et~al.}(2020)\citenamefont {König},
  \citenamefont {Yao}, \citenamefont {Puthen-Veettil},\ and\ \citenamefont
  {Smith}}]{konig_non-equilibrium_2020}%
  \BibitemOpen
  \bibfield  {author} {\bibinfo {author} {\bibfnamefont {D.}~\bibnamefont
  {König}}, \bibinfo {author} {\bibfnamefont {Y.}~\bibnamefont {Yao}},
  \bibinfo {author} {\bibfnamefont {B.}~\bibnamefont {Puthen-Veettil}},\ and\
  \bibinfo {author} {\bibfnamefont {S.~C.}\ \bibnamefont {Smith}},\ }\bibfield
  {title} {\bibinfo {title} {Non-equilibrium dynamics, materials and structures
  for hot carrier solar cells: a detailed review},\ }\href
  {https://doi.org/10.1088/1361-6641/ab8171} {\bibfield  {journal} {\bibinfo
  {journal} {Semiconductor Science and Technology}\ }\textbf {\bibinfo {volume}
  {35}},\ \bibinfo {pages} {073002} (\bibinfo {year} {2020})}\BibitemShut
  {NoStop}%
\bibitem [{\citenamefont {Zhang}\ \emph {et~al.}(2021)\citenamefont {Zhang},
  \citenamefont {Jia}, \citenamefont {Liu}, \citenamefont {Zhang},
  \citenamefont {Lin}, \citenamefont {Zhang},\ and\ \citenamefont
  {Conibeer}}]{zhang_review_2021}%
  \BibitemOpen
  \bibfield  {author} {\bibinfo {author} {\bibfnamefont {Y.}~\bibnamefont
  {Zhang}}, \bibinfo {author} {\bibfnamefont {X.}~\bibnamefont {Jia}}, \bibinfo
  {author} {\bibfnamefont {S.}~\bibnamefont {Liu}}, \bibinfo {author}
  {\bibfnamefont {B.}~\bibnamefont {Zhang}}, \bibinfo {author} {\bibfnamefont
  {K.}~\bibnamefont {Lin}}, \bibinfo {author} {\bibfnamefont {J.}~\bibnamefont
  {Zhang}},\ and\ \bibinfo {author} {\bibfnamefont {G.}~\bibnamefont
  {Conibeer}},\ }\bibfield  {title} {\bibinfo {title} {A review on
  thermalization mechanisms and prospect absorber materials for the hot carrier
  solar cells},\ }\href {https://doi.org/10.1016/j.solmat.2021.111073}
  {\bibfield  {journal} {\bibinfo  {journal} {Solar Energy Materials and Solar
  Cells}\ }\textbf {\bibinfo {volume} {225}},\ \bibinfo {pages} {111073}
  (\bibinfo {year} {2021})}\BibitemShut {NoStop}%
\bibitem [{\citenamefont {Dimmock}\ \emph {et~al.}(2014)\citenamefont
  {Dimmock}, \citenamefont {Day}, \citenamefont {Kauer}, \citenamefont
  {Smith},\ and\ \citenamefont {Heffernan}}]{dimmock_demonstration_2014}%
  \BibitemOpen
  \bibfield  {author} {\bibinfo {author} {\bibfnamefont {J.~A.~R.}\
  \bibnamefont {Dimmock}}, \bibinfo {author} {\bibfnamefont {S.}~\bibnamefont
  {Day}}, \bibinfo {author} {\bibfnamefont {M.}~\bibnamefont {Kauer}}, \bibinfo
  {author} {\bibfnamefont {K.}~\bibnamefont {Smith}},\ and\ \bibinfo {author}
  {\bibfnamefont {J.}~\bibnamefont {Heffernan}},\ }\bibfield  {title} {\bibinfo
  {title} {Demonstration of a hot-carrier photovoltaic cell: {Demonstration} of
  a hot-carrier photovoltaic cell},\ }\href {https://doi.org/10.1002/pip.2444}
  {\bibfield  {journal} {\bibinfo  {journal} {Progress in Photovoltaics:
  Research and Applications}\ }\textbf {\bibinfo {volume} {22}},\ \bibinfo
  {pages} {151} (\bibinfo {year} {2014})}\BibitemShut {NoStop}%
\bibitem [{\citenamefont {Dimmock}\ \emph {et~al.}(2019)\citenamefont
  {Dimmock}, \citenamefont {Kauer}, \citenamefont {Wu}, \citenamefont {Liu},
  \citenamefont {Stavrinou},\ and\ \citenamefont
  {Ekins-Daukes}}]{dimmock_metallic_2019}%
  \BibitemOpen
  \bibfield  {author} {\bibinfo {author} {\bibfnamefont {J.~A.~R.}\
  \bibnamefont {Dimmock}}, \bibinfo {author} {\bibfnamefont {M.}~\bibnamefont
  {Kauer}}, \bibinfo {author} {\bibfnamefont {J.}~\bibnamefont {Wu}}, \bibinfo
  {author} {\bibfnamefont {H.}~\bibnamefont {Liu}}, \bibinfo {author}
  {\bibfnamefont {P.~N.}\ \bibnamefont {Stavrinou}},\ and\ \bibinfo {author}
  {\bibfnamefont {N.~J.}\ \bibnamefont {Ekins-Daukes}},\ }\bibfield  {title}
  {\bibinfo {title} {A metallic hot-carrier photovoltaic device},\ }\href
  {https://doi.org/10.1088/1361-6641/ab1222} {\bibfield  {journal} {\bibinfo
  {journal} {Semiconductor Science and Technology}\ }\textbf {\bibinfo {volume}
  {34}},\ \bibinfo {pages} {064001} (\bibinfo {year} {2019})},\ \bibinfo {note}
  {publisher: IOP Publishing}\BibitemShut {NoStop}%
\bibitem [{\citenamefont {Nguyen}\ \emph {et~al.}(2018)\citenamefont {Nguyen},
  \citenamefont {Lombez}, \citenamefont {Gibelli}, \citenamefont
  {Boyer-Richard}, \citenamefont {Le~Corre}, \citenamefont {Durand},\ and\
  \citenamefont {Guillemoles}}]{nguyen_quantitative_2018}%
  \BibitemOpen
  \bibfield  {author} {\bibinfo {author} {\bibfnamefont {D.-T.}\ \bibnamefont
  {Nguyen}}, \bibinfo {author} {\bibfnamefont {L.}~\bibnamefont {Lombez}},
  \bibinfo {author} {\bibfnamefont {F.}~\bibnamefont {Gibelli}}, \bibinfo
  {author} {\bibfnamefont {S.}~\bibnamefont {Boyer-Richard}}, \bibinfo {author}
  {\bibfnamefont {A.}~\bibnamefont {Le~Corre}}, \bibinfo {author}
  {\bibfnamefont {O.}~\bibnamefont {Durand}},\ and\ \bibinfo {author}
  {\bibfnamefont {J.-F.}\ \bibnamefont {Guillemoles}},\ }\bibfield  {title}
  {\bibinfo {title} {Quantitative experimental assessment of hot
  carrier-enhanced solar cells at room temperature},\ }\href
  {https://doi.org/10.1038/s41560-018-0106-3} {\bibfield  {journal} {\bibinfo
  {journal} {Nature Energy}\ }\textbf {\bibinfo {volume} {3}},\ \bibinfo
  {pages} {236} (\bibinfo {year} {2018})}\BibitemShut {NoStop}%
\bibitem [{\citenamefont {Li}\ \emph {et~al.}(2017)\citenamefont {Li},
  \citenamefont {Bhaumik}, \citenamefont {Goh}, \citenamefont {Kumar},
  \citenamefont {Yantara}, \citenamefont {Grätzel}, \citenamefont
  {Mhaisalkar}, \citenamefont {Mathews},\ and\ \citenamefont
  {Sum}}]{li_slow_2017}%
  \BibitemOpen
  \bibfield  {author} {\bibinfo {author} {\bibfnamefont {M.}~\bibnamefont
  {Li}}, \bibinfo {author} {\bibfnamefont {S.}~\bibnamefont {Bhaumik}},
  \bibinfo {author} {\bibfnamefont {T.}~\bibnamefont {Goh}}, \bibinfo {author}
  {\bibfnamefont {M.}~\bibnamefont {Kumar}}, \bibinfo {author} {\bibfnamefont
  {N.}~\bibnamefont {Yantara}}, \bibinfo {author} {\bibfnamefont
  {M.}~\bibnamefont {Grätzel}}, \bibinfo {author} {\bibfnamefont
  {S.}~\bibnamefont {Mhaisalkar}}, \bibinfo {author} {\bibfnamefont
  {N.}~\bibnamefont {Mathews}},\ and\ \bibinfo {author} {\bibfnamefont
  {T.}~\bibnamefont {Sum}},\ }\bibfield  {title} {\bibinfo {title} {Slow
  cooling and highly efficient extraction of hot carriers in colloidal
  perovskite nanocrystals},\ }\href@noop {} {\bibfield  {journal} {\bibinfo
  {journal} {Nature Communications}\ }\textbf {\bibinfo {volume} {8}},\
  \bibinfo {pages} {14350} (\bibinfo {year} {2017})}\BibitemShut {NoStop}%
\bibitem [{\citenamefont {Esmaielpour}\ \emph {et~al.}(2020)\citenamefont
  {Esmaielpour}, \citenamefont {Suchet}, \citenamefont {Lombez}, \citenamefont
  {Delamarre}, \citenamefont {Boyer-Richard}, \citenamefont {Beck},
  \citenamefont {Le~Corre}, \citenamefont {Durand},\ and\ \citenamefont
  {Guillemoles}}]{esmaielpour_determination_2020}%
  \BibitemOpen
  \bibfield  {author} {\bibinfo {author} {\bibfnamefont {H.}~\bibnamefont
  {Esmaielpour}}, \bibinfo {author} {\bibfnamefont {D.}~\bibnamefont {Suchet}},
  \bibinfo {author} {\bibfnamefont {L.}~\bibnamefont {Lombez}}, \bibinfo
  {author} {\bibfnamefont {A.}~\bibnamefont {Delamarre}}, \bibinfo {author}
  {\bibfnamefont {S.}~\bibnamefont {Boyer-Richard}}, \bibinfo {author}
  {\bibfnamefont {A.}~\bibnamefont {Beck}}, \bibinfo {author} {\bibfnamefont
  {A.}~\bibnamefont {Le~Corre}}, \bibinfo {author} {\bibfnamefont
  {O.}~\bibnamefont {Durand}},\ and\ \bibinfo {author} {\bibfnamefont {J.-F.}\
  \bibnamefont {Guillemoles}},\ }\bibfield  {title} {\bibinfo {title}
  {Determination of photo-induced {Seebeck} coefficient for hot carrier solar
  cell applications},\ }in\ \href
  {https://doi.org/10.1109/PVSC45281.2020.9300544} {\emph {\bibinfo {booktitle}
  {2020 47th {IEEE} {Photovoltaic} {Specialists} {Conference} ({PVSC})}}}\
  (\bibinfo  {publisher} {IEEE},\ \bibinfo {address} {Calgary, AB, Canada},\
  \bibinfo {year} {2020})\ pp.\ \bibinfo {pages} {0747--0751}\BibitemShut
  {NoStop}%
\bibitem [{\citenamefont {Williams}\ \emph {et~al.}(2013)\citenamefont
  {Williams}, \citenamefont {Nelson}, \citenamefont {Yan}, \citenamefont {Li},\
  and\ \citenamefont {Zhu}}]{williams_hot_2013}%
  \BibitemOpen
  \bibfield  {author} {\bibinfo {author} {\bibfnamefont {K.}~\bibnamefont
  {Williams}}, \bibinfo {author} {\bibfnamefont {C.}~\bibnamefont {Nelson}},
  \bibinfo {author} {\bibfnamefont {X.}~\bibnamefont {Yan}}, \bibinfo {author}
  {\bibfnamefont {L.-S.}\ \bibnamefont {Li}},\ and\ \bibinfo {author}
  {\bibfnamefont {X.}~\bibnamefont {Zhu}},\ }\bibfield  {title} {\bibinfo
  {title} {Hot electron injection from graphene quantum dots to {TiO2}},\
  }\href {https://doi.org/10.1021/nn305080c} {\bibfield  {journal} {\bibinfo
  {journal} {ACS Nano}\ }\textbf {\bibinfo {volume} {7}},\ \bibinfo {pages}
  {1388} (\bibinfo {year} {2013})}\BibitemShut {NoStop}%
\bibitem [{\citenamefont {Mart{\'{i}}}(2019)}]{Marti2019}%
  \BibitemOpen
  \bibfield  {author} {\bibinfo {author} {\bibfnamefont {A.}~\bibnamefont
  {Mart{\'{i}}}},\ }\bibfield  {title} {\bibinfo {title} {{From the Hot Carrier
  Solar Cell to the Intermediate Band Solar Cell, Passing through the
  Multiple-Exciton Generation Solar Cell and then back to the Hot Carrier Solar
  Cell: the Dance of the Electro-Chemical Potentials}},\ }in\ \href
  {https://doi.org/10.4229/EUPVSEC20192019-1AO.1.1} {\emph {\bibinfo
  {booktitle} {36th European Photovoltaics Specialist Conference and
  Exhibition}}}\ (\bibinfo {year} {2019})\ pp.\ \bibinfo {pages}
  {6--12}\BibitemShut {NoStop}%
\bibitem [{\citenamefont {Marti}\ and\ \citenamefont
  {Luque}(2013)}]{marti_ele_2013}%
  \BibitemOpen
  \bibfield  {author} {\bibinfo {author} {\bibfnamefont {A.}~\bibnamefont
  {Marti}}\ and\ \bibinfo {author} {\bibfnamefont {A.}~\bibnamefont {Luque}},\
  }\bibfield  {title} {\bibinfo {title} {Electrochemical {Potentials}
  ({Quasi}-{Fermi} {Levels}) and the {Operation} of {Hot}-{Carrier},
  {Impact}-{Ionization}, and {Intermediate}-{Band} {Solar} {Cells}},\ }\href
  {https://doi.org/10.1109/JPHOTOV.2013.2274381} {\bibfield  {journal}
  {\bibinfo  {journal} {IEEE Journal of Photovoltaics}\ }\textbf {\bibinfo
  {volume} {3}},\ \bibinfo {pages} {1298} (\bibinfo {year} {2013})}\BibitemShut
  {NoStop}%
\bibitem [{\citenamefont {Weldford}\ and\ \citenamefont
  {Winston}(1989)}]{WELFORD19899}%
  \BibitemOpen
  \bibfield  {author} {\bibinfo {author} {\bibfnamefont {W.}~\bibnamefont
  {Weldford}}\ and\ \bibinfo {author} {\bibfnamefont {R.}~\bibnamefont
  {Winston}},\ }\bibfield  {title} {\bibinfo {title} {Chapter 2 - some basic
  ideas in geometrical optics},\ }in\ \href
  {https://doi.org/https://doi.org/10.1016/B978-0-12-742885-7.50005-X} {\emph
  {\bibinfo {booktitle} {High Collection Nonimaging Optics}}},\ \bibinfo
  {editor} {edited by\ \bibinfo {editor} {\bibfnamefont {W.}~\bibnamefont
  {Weldford}}\ and\ \bibinfo {editor} {\bibfnamefont {R.}~\bibnamefont
  {Winston}}}\ (\bibinfo  {publisher} {Academic Press},\ \bibinfo {year}
  {1989})\ pp.\ \bibinfo {pages} {9 -- 29}\BibitemShut {NoStop}%
\bibitem [{\citenamefont {W{\"{u}}rfel}\ \emph {et~al.}(2005)\citenamefont
  {W{\"{u}}rfel}, \citenamefont {Brown}, \citenamefont {Humphrey},\ and\
  \citenamefont {Green}}]{Wurfel2005}%
  \BibitemOpen
  \bibfield  {author} {\bibinfo {author} {\bibfnamefont {P.}~\bibnamefont
  {W{\"{u}}rfel}}, \bibinfo {author} {\bibfnamefont {A.~S.}\ \bibnamefont
  {Brown}}, \bibinfo {author} {\bibfnamefont {T.~E.}\ \bibnamefont
  {Humphrey}},\ and\ \bibinfo {author} {\bibfnamefont {M.~A.}\ \bibnamefont
  {Green}},\ }\bibfield  {title} {\bibinfo {title} {{Particle conservation in
  the hot-carrier solar cell}},\ }\href {https://doi.org/10.1002/pip.584}
  {\bibfield  {journal} {\bibinfo  {journal} {Progress in Photovoltaics:
  Research and Applications}\ }\textbf {\bibinfo {volume} {13}},\ \bibinfo
  {pages} {277} (\bibinfo {year} {2005})}\BibitemShut {NoStop}%
\bibitem [{\citenamefont {Callen}(1981{\natexlab{a}})}]{Callen1}%
  \BibitemOpen
  \bibfield  {author} {\bibinfo {author} {\bibfnamefont {H.}~\bibnamefont
  {Callen}},\ }\bibfield  {title} {\bibinfo {title} {Efectos termoeléctricos y
  termomagnéticos},\ }in\ \href@noop {} {\emph {\bibinfo {booktitle}
  {Termodinámica}}}\ (\bibinfo  {publisher} {Editorial AC},\ \bibinfo
  {address} {Madrid},\ \bibinfo {year} {1981})\ \bibinfo {edition} {1st}\ ed.,\
  pp.\ \bibinfo {pages} {287--290}\BibitemShut {NoStop}%
\bibitem [{\citenamefont {Landsberg}\ and\ \citenamefont
  {Tonge}()}]{landsberg}%
  \BibitemOpen
  \bibfield  {author} {\bibinfo {author} {\bibfnamefont {P.~T.}\ \bibnamefont
  {Landsberg}}\ and\ \bibinfo {author} {\bibfnamefont {G.}~\bibnamefont
  {Tonge}},\ }\bibfield  {title} {\bibinfo {title} {Thermodynamic energy
  conversion efficiencies},\ }\href@noop {} {\bibfield  {journal} {\bibinfo
  {journal} {Journal of Applied Physics}\ }\textbf {\bibinfo {volume} {51}},\
  \bibinfo {pages} {R1}}\BibitemShut {NoStop}%
\bibitem [{\citenamefont {Datta}(2005)}]{Datta}%
  \BibitemOpen
  \bibfield  {author} {\bibinfo {author} {\bibfnamefont {S.}~\bibnamefont
  {Datta}},\ }\bibfield  {title} {\bibinfo {title} {Prologue: an atomistic view
  of electrical resistance},\ }in\ \href@noop {} {\emph {\bibinfo {booktitle}
  {Quantum Transport: atom to transistor}}}\ (\bibinfo  {publisher} {Cambridge
  University Press},\ \bibinfo {address} {Cambridge},\ \bibinfo {year} {2005})\
  Chap.~\bibinfo {chapter} {1}\BibitemShut {NoStop}%
\bibitem [{\citenamefont {Landauer}(1957)}]{landauer_1957}%
  \BibitemOpen
  \bibfield  {author} {\bibinfo {author} {\bibfnamefont {R.}~\bibnamefont
  {Landauer}},\ }\bibfield  {title} {\bibinfo {title} {Spatial {Variation} of
  {Currents} and {Fields} {Due} to {Localized} {Scatterers} in {Metallic}
  {Conduction}},\ }\href@noop {} {\bibfield  {journal} {\bibinfo  {journal}
  {IBM Journal of Research and Development}\ }\textbf {\bibinfo {volume} {1}},\
  \bibinfo {pages} {223} (\bibinfo {year} {1957})},\ \bibinfo {note}
  {conference Name: IBM Journal of Research and Development}\BibitemShut
  {NoStop}%
\bibitem [{\citenamefont {Landauer}(1992)}]{landauer_1992}%
  \BibitemOpen
  \bibfield  {author} {\bibinfo {author} {\bibfnamefont {R.}~\bibnamefont
  {Landauer}},\ }\bibfield  {title} {\bibinfo {title} {Conductance from
  transmission: common sense points},\ }\href@noop {} {\bibfield  {journal}
  {\bibinfo  {journal} {Physica Scripta}\ }\textbf {\bibinfo {volume} {T42}},\
  \bibinfo {pages} {110} (\bibinfo {year} {1992})},\ \bibinfo {note}
  {publisher: IOP Publishing}\BibitemShut {NoStop}%
\bibitem [{\citenamefont {Limpert}\ and\ \citenamefont
  {Bremner}(2015)}]{limpert_hot_2015}%
  \BibitemOpen
  \bibfield  {author} {\bibinfo {author} {\bibfnamefont {S.~C.}\ \bibnamefont
  {Limpert}}\ and\ \bibinfo {author} {\bibfnamefont {S.~P.}\ \bibnamefont
  {Bremner}},\ }\bibfield  {title} {\bibinfo {title} {Hot carrier extraction
  using energy selective contacts and its impact on the limiting efficiency of
  a hot carrier solar cell},\ }\href {https://doi.org/10.1063/1.4928750}
  {\bibfield  {journal} {\bibinfo  {journal} {Applied Physics Letters}\
  }\textbf {\bibinfo {volume} {107}},\ \bibinfo {pages} {073902} (\bibinfo
  {year} {2015})},\ \bibinfo {note} {publisher: American Institute of
  Physics}\BibitemShut {NoStop}%
\bibitem [{\citenamefont {Bass}(1983)}]{Landolt_r}%
  \BibitemOpen
  \bibfield  {author} {\bibinfo {author} {\bibfnamefont {J.}~\bibnamefont
  {Bass}},\ }\href {https://doi.org/10.1007/10307022_11} {\bibinfo {title}
  {Electrical {Resistivity}, {Kondo} and {Spin} {Fluctuation} {Systems}, {Spin}
  {Glasses} and {Thermopower} {\textperiodcentered} {Ac} - {Gd}: {Datasheet}
  from {Landolt}-{B{\"o}rnstein} - {Group} {III} {Condensed} {Matter}
  {\textperiodcentered} {Volume} {15A} in {SpringerMaterials} (1983)}}
  (\bibinfo {year} {1983}),\ \bibinfo {note} {copyright 1983 Springer-Verlag
  Berlin Heidelberg}\BibitemShut {NoStop}%
\bibitem [{\citenamefont {Ohta}\ \emph {et~al.}(2007)\citenamefont {Ohta},
  \citenamefont {Kim}, \citenamefont {Mune}, \citenamefont {Mizoguchi},
  \citenamefont {Nomura}, \citenamefont {Ohta}, \citenamefont {Nomura},
  \citenamefont {Nakanishi}, \citenamefont {Ikuhara}, \citenamefont {Hirano},
  \citenamefont {Hosono},\ and\ \citenamefont {Koumoto}}]{Giant_2007}%
  \BibitemOpen
  \bibfield  {author} {\bibinfo {author} {\bibfnamefont {H.}~\bibnamefont
  {Ohta}}, \bibinfo {author} {\bibfnamefont {S.}~\bibnamefont {Kim}}, \bibinfo
  {author} {\bibfnamefont {Y.}~\bibnamefont {Mune}}, \bibinfo {author}
  {\bibfnamefont {T.}~\bibnamefont {Mizoguchi}}, \bibinfo {author}
  {\bibfnamefont {K.}~\bibnamefont {Nomura}}, \bibinfo {author} {\bibfnamefont
  {S.}~\bibnamefont {Ohta}}, \bibinfo {author} {\bibfnamefont {T.}~\bibnamefont
  {Nomura}}, \bibinfo {author} {\bibfnamefont {Y.}~\bibnamefont {Nakanishi}},
  \bibinfo {author} {\bibfnamefont {Y.}~\bibnamefont {Ikuhara}}, \bibinfo
  {author} {\bibfnamefont {M.}~\bibnamefont {Hirano}}, \bibinfo {author}
  {\bibfnamefont {H.}~\bibnamefont {Hosono}},\ and\ \bibinfo {author}
  {\bibfnamefont {K.}~\bibnamefont {Koumoto}},\ }\bibfield  {title} {\bibinfo
  {title} {Giant thermoelectric {Seebeck} coefficient of a two-dimensional
  electron gas in {SrTiO3}},\ }\href {https://doi.org/10.1038/nmat1821}
  {\bibfield  {journal} {\bibinfo  {journal} {Nature Materials}\ }\textbf
  {\bibinfo {volume} {6}},\ \bibinfo {pages} {129} (\bibinfo {year}
  {2007})}\BibitemShut {NoStop}%
\bibitem [{\citenamefont {Goto}\ \emph {et~al.}(1997)\citenamefont {Goto},
  \citenamefont {Li}, \citenamefont {Hirai}, \citenamefont {Maeda},
  \citenamefont {Kato},\ and\ \citenamefont {Maesono}}]{Goto}%
  \BibitemOpen
  \bibfield  {author} {\bibinfo {author} {\bibfnamefont {T.}~\bibnamefont
  {Goto}}, \bibinfo {author} {\bibfnamefont {J.~H.}\ \bibnamefont {Li}},
  \bibinfo {author} {\bibfnamefont {T.}~\bibnamefont {Hirai}}, \bibinfo
  {author} {\bibfnamefont {Y.}~\bibnamefont {Maeda}}, \bibinfo {author}
  {\bibfnamefont {R.}~\bibnamefont {Kato}},\ and\ \bibinfo {author}
  {\bibfnamefont {A.}~\bibnamefont {Maesono}},\ }\bibfield  {title} {\bibinfo
  {title} {Measurements of the seebeck coefficient of thermoelectric materials
  by an ac method},\ }\href@noop {} {\bibfield  {journal} {\bibinfo  {journal}
  {Int J Thermophys}\ }\textbf {\bibinfo {volume} {18}},\ \bibinfo {pages}
  {569} (\bibinfo {year} {1997})}\BibitemShut {NoStop}%
\bibitem [{\citenamefont {Zhao}\ \emph {et~al.}(2014)\citenamefont {Zhao},
  \citenamefont {Dravid},\ and\ \citenamefont
  {Kanatzidis}}]{zhao_panoscopic_2014}%
  \BibitemOpen
  \bibfield  {author} {\bibinfo {author} {\bibfnamefont {L.-D.}\ \bibnamefont
  {Zhao}}, \bibinfo {author} {\bibfnamefont {V.~P.}\ \bibnamefont {Dravid}},\
  and\ \bibinfo {author} {\bibfnamefont {M.~G.}\ \bibnamefont {Kanatzidis}},\
  }\bibfield  {title} {\bibinfo {title} {The panoscopic approach to high
  performance thermoelectrics},\ }\href {https://doi.org/10.1039/C3EE43099E}
  {\bibfield  {journal} {\bibinfo  {journal} {Energy Environ. Sci.}\ }\textbf
  {\bibinfo {volume} {7}},\ \bibinfo {pages} {251} (\bibinfo {year}
  {2014})}\BibitemShut {NoStop}%
\bibitem [{\citenamefont {Mahan}\ and\ \citenamefont {Sofo}(1996)}]{mahan}%
  \BibitemOpen
  \bibfield  {author} {\bibinfo {author} {\bibfnamefont {G.~D.}\ \bibnamefont
  {Mahan}}\ and\ \bibinfo {author} {\bibfnamefont {J.~O.}\ \bibnamefont
  {Sofo}},\ }\bibfield  {title} {\bibinfo {title} {The best thermoelectric.},\
  }\href@noop {} {\bibfield  {journal} {\bibinfo  {journal} {Proceedings of the
  National Academy of Sciences}\ }\textbf {\bibinfo {volume} {93}},\ \bibinfo
  {pages} {7436} (\bibinfo {year} {1996})}\BibitemShut {NoStop}%
\bibitem [{\citenamefont {Jeong}\ \emph {et~al.}(2012)\citenamefont {Jeong},
  \citenamefont {Kim},\ and\ \citenamefont {Lundstrom}}]{jeong_best_2012}%
  \BibitemOpen
  \bibfield  {author} {\bibinfo {author} {\bibfnamefont {C.}~\bibnamefont
  {Jeong}}, \bibinfo {author} {\bibfnamefont {R.}~\bibnamefont {Kim}},\ and\
  \bibinfo {author} {\bibfnamefont {M.~S.}\ \bibnamefont {Lundstrom}},\
  }\bibfield  {title} {\bibinfo {title} {On the best bandstructure for
  thermoelectric performance: A landauer perspective},\ }\href@noop {}
  {\bibfield  {journal} {\bibinfo  {journal} {Journal of Applied Physics}\
  }\textbf {\bibinfo {volume} {111}},\ \bibinfo {pages} {113707} (\bibinfo
  {year} {2012})}\BibitemShut {NoStop}%
\bibitem [{\citenamefont {Mehdizadeh~Dehkordi}\ \emph
  {et~al.}(2015)\citenamefont {Mehdizadeh~Dehkordi}, \citenamefont {Zebarjadi},
  \citenamefont {He},\ and\ \citenamefont
  {Tritt}}]{mehdizadeh_dehkordi_thermoelectric_2015}%
  \BibitemOpen
  \bibfield  {author} {\bibinfo {author} {\bibfnamefont {A.}~\bibnamefont
  {Mehdizadeh~Dehkordi}}, \bibinfo {author} {\bibfnamefont {M.}~\bibnamefont
  {Zebarjadi}}, \bibinfo {author} {\bibfnamefont {J.}~\bibnamefont {He}},\ and\
  \bibinfo {author} {\bibfnamefont {T.~M.}\ \bibnamefont {Tritt}},\ }\bibfield
  {title} {\bibinfo {title} {Thermoelectric power factor: Enhancement
  mechanisms and strategies for higher performance thermoelectric materials},\
  }\href@noop {} {\bibfield  {journal} {\bibinfo  {journal} {Materials Science
  and Engineering: R: Reports}\ }\textbf {\bibinfo {volume} {97}},\ \bibinfo
  {pages} {1} (\bibinfo {year} {2015})}\BibitemShut {NoStop}%
\bibitem [{\citenamefont {Nakpathomkun}\ \emph {et~al.}(2010)\citenamefont
  {Nakpathomkun}, \citenamefont {Xu},\ and\ \citenamefont {Linke}}]{nak_2010}%
  \BibitemOpen
  \bibfield  {author} {\bibinfo {author} {\bibfnamefont {N.}~\bibnamefont
  {Nakpathomkun}}, \bibinfo {author} {\bibfnamefont {H.~Q.}\ \bibnamefont
  {Xu}},\ and\ \bibinfo {author} {\bibfnamefont {H.}~\bibnamefont {Linke}},\
  }\bibfield  {title} {\bibinfo {title} {Thermoelectric efficiency at maximum
  power in low-dimensional systems},\ }\href@noop {} {\bibfield  {journal}
  {\bibinfo  {journal} {Physical Review B}\ }\textbf {\bibinfo {volume} {82}},\
  \bibinfo {pages} {235428} (\bibinfo {year} {2010})},\ \bibinfo {note}
  {publisher: American Physical Society}\BibitemShut {NoStop}%
\bibitem [{\citenamefont {Luque}\ and\ \citenamefont
  {Martí}(2003)}]{Luque_Handbook}%
  \BibitemOpen
  \bibfield  {author} {\bibinfo {author} {\bibfnamefont {A.}~\bibnamefont
  {Luque}}\ and\ \bibinfo {author} {\bibfnamefont {A.}~\bibnamefont {Martí}},\
  }\bibfield  {title} {\bibinfo {title} {Theoretical limits of photovoltacis
  conversion and new generation solar cells},\ }in\ \href@noop {} {\emph
  {\bibinfo {booktitle} {Handbook of photovoltaic science and engineering}}},\
  \bibinfo {editor} {edited by\ \bibinfo {editor} {\bibfnamefont
  {A.}~\bibnamefont {Luque}}\ and\ \bibinfo {editor} {\bibfnamefont
  {S.}~\bibnamefont {Hegedus}}}\ (\bibinfo  {publisher} {Wiley},\ \bibinfo
  {address} {Chichester},\ \bibinfo {year} {2003})\ Chap.~\bibinfo {chapter}
  {4}, p.\ \bibinfo {pages} {132}\BibitemShut {NoStop}%
\bibitem [{\citenamefont {Callen}(1981{\natexlab{b}})}]{Callen2}%
  \BibitemOpen
  \bibfield  {author} {\bibinfo {author} {\bibfnamefont {H.}~\bibnamefont
  {Callen}},\ }\bibfield  {title} {\bibinfo {title} {Termodinámica
  irreversible},\ }in\ \href@noop {} {\emph {\bibinfo {booktitle}
  {Termodinámica}}}\ (\bibinfo  {publisher} {Editorial AC},\ \bibinfo
  {address} {Madrid},\ \bibinfo {year} {1981})\ \bibinfo {edition} {1st}\ ed.,\
  pp.\ \bibinfo {pages} {282--283}\BibitemShut {NoStop}%
\end{thebibliography}%

\end{document}